\documentclass[12pt,onecolumn,journal]{IEEEtran}
\usepackage{latexsym}
\usepackage[margin=2.1cm, lmargin=2.1cm]{geometry}
\usepackage{float}
\usepackage{amsfonts}
\usepackage{amsbsy}
\usepackage{amssymb}
\usepackage{times}
\usepackage{graphicx}
\usepackage{enumerate}
\usepackage[usenames]{color}
\usepackage[dvips]{pstcol}
\usepackage{epstopdf}
\usepackage{cite}
\usepackage{amssymb}
\usepackage{amsfonts}
\usepackage{graphicx}
\usepackage{epsfig}
\usepackage{psfrag}
\usepackage{xcolor}
\usepackage{amsfonts, bm}
\usepackage{epstopdf}
\usepackage{cite}
\usepackage{color}
\usepackage{xcolor}
\usepackage{subfig}
\usepackage{verbatim}
\usepackage{multirow}
\usepackage{array}
\usepackage{booktabs}
\usepackage{amsthm}

\usepackage{algorithm}
\usepackage{algorithmicx}
\usepackage{algpseudocode}
\usepackage{amsmath}

\newcommand{\tabincell}[2]{\begin{tabular}{@{}#1@{}}#2\end{tabular}}
\IEEEoverridecommandlockouts
\begin{document}
\title{ \LARGE{ 
	Joint Deep Reinforcement Learning and Unfolding: Beam Selection and Precoding for mmWave Multiuser MIMO with Lens Arrays
} }
\author{ Qiyu Hu, Yanzhen Liu, Yunlong Cai, Guanding Yu, and Zhi Ding
	\thanks{
		Q. Hu, Y. Liu, Y. Cai, and G. Yu are with the College of Information Science and Electronic Engineering, Zhejiang University, Hangzhou 310027, China (e-mail: qiyhu@zju.edu.cn; yanzliu@zju.edu.cn; ylcai@zju.edu.cn; yuguanding@zju.edu.cn).
		Z. Ding is with the Department of Electrical and Computer Engineering, University of California, Davis, CA 95616, USA (e-mail: zding@ucdavis.edu).
	}
}

\maketitle
\vspace{-3.3em}
\begin{abstract}
The millimeter wave (mmWave) multiuser multiple-input multiple-output (MU-MIMO) systems with discrete lens arrays (DLA) have received great attention due to their simple hardware implementation and excellent performance. 
In this work, we investigate the joint design of beam selection and digital precoding matrices for mmWave MU-MIMO systems with DLA to maximize the sum-rate subject to the transmit power constraint and the constraints of the selection matrix structure. 
The investigated non-convex problem with discrete variables and coupled constraints is challenging to solve and an efficient framework of joint neural network (NN) design is proposed to tackle it. Specifically, the proposed framework consists of a deep reinforcement learning (DRL)-based NN and a deep-unfolding NN, which are employed to optimize the beam selection and digital precoding matrices, respectively.
As for the DRL-based NN, we formulate the beam selection problem as a Markov decision process and a double deep Q-network algorithm is developed to solve it.
The base station is considered to be an agent, where the state, action, and reward function are carefully designed. 
Regarding the design of the digital precoding matrix, we develop an iterative weighted minimum mean-square error algorithm induced deep-unfolding NN, which unfolds this algorithm into a layer-wise structure with introduced trainable parameters.
Simulation results verify that this jointly trained NN remarkably outperforms the existing iterative algorithms with reduced complexity and stronger robustness. 
\end{abstract}

\IEEEpeerreviewmaketitle

\section{Introduction}
Recently, the millimeter wave (mmWave) communication has received great attention for the design of wireless systems due to the benefits of mitigating the spectrum shortage and supporting high data rates \cite{Magatech}. The short wavelength of mmWave makes it practically feasible to equip a large number of antennas at the transceiver in a cellular network, where the massive multiple-input multiple-output (MIMO) techniques can be readily realized to perform highly directional transmissions \cite{MIMO1,MIMO2,MIMO3}. Nevertheless, the conventional precoding relied on fully digital (FD) processing structures generally leads to an unaffordable cost of radio frequency (RF) chains and high power consumption in massive MIMO scenarios. The hybrid analog-digital (AD) transceiver consisting of cascaded baseband digital precoder and RF analog beamformer has been proposed as a trade-off between cost and performance, which employs less RF chains than the number of antennas, while imposing a constant modulus constraint on the analog beamforming matrix \cite{Hybrid}. Moreover, note that the use of phase shifters still causes large energy consumption.

To address these issues, a promising mmWave MIMO technique with discrete lens arrays (DLA) has been proposed in \cite{Concept} to reduce hardware implementation complexity without remarkable performance loss \cite{RZhang,NPhard,NearOptim,RGuo,GCBeam,LowRF,Hanzo,Heath,LowComp}.
A DLA generally consists of two parts: an electromagnetic lens and a matching antenna array. It transforms the conventional MIMO spatial channels into beamspace channels with angle-dependent energy-focusing capabilities \cite{RZhang}. 
Practically, due to the sparsity of beamspace channels, only a small number of beams with large gains are selected. In addition, each RF chain selects a single beam, thus the required number of RF chains can be dramatically reduced. Moreover, the conventional phase shifters-enabled hybrid precoding structure can be substituted by a switching network, which greatly reduces the hardware costs and cuts down the power budget.
In spite of these superiorities of the DLA, it has been proved in \cite{NPhard} that the beam selection problem is NP-hard and how to handle this problem efficiently remains an open issue. In \cite{NearOptim}, the interference-aware algorithm has been proposed to solve this NP-hard problem heuristically. A penalty dual decompsition (PDD) algorithm has been developed in \cite{RGuo} for jointly optimizing the beam selection and digital precoding matrices. The beam selection problem for the high-dimensional multiuser (MU) case has been investigated in \cite{GCBeam}. In \cite{LowRF}, several algorithms have been proposed, e.g., maximum magnitude beam selection and maximum signal-to-interference-ratio (SINR) beam selection. The authors in \cite{Hanzo} have developed an energy-max algorithm for beam selection and a successive interference cancellation based precoding method in massive MIMO systems.

However, the optimization techniques usually provide very high computational complexity and are easily trapped into a local optimum when dealing with a non-convex problem. In addition, the existing algorithms mainly optimize the beam selection matrix and the digital precoding matrix separately. Thus, the joint design is worth studying for further improving the performance, which is more challenging and belongs to the mixed-integer non-linear programming (MINLP) in wireless communications. Recently, the machine learning techniques emerge as an efficient method and there is a number of studies employing the neural networks (NNs) for solving the investigated problems \cite{DLMaga}, such as precoding design \cite{LearnOpt}, beam selection \cite{SVMBeam,5GBeam,MLPBeam}, and resource allocation \cite{DRLresou1}. The idea is to regard the iterative algorithm as a black-box, and employ the NNs to learn the relationship between the inputs and outputs.
The authors in \cite{LearnOpt} have applied a NN with fully-connected (FC) layers to approximate the iterative weighted minimum mean-square error (WMMSE) algorithm for precoding. In \cite{SVMBeam}, the beam selection problem has been regarded as a multiclass-classification problem solved by the support vector machine. The authors in \cite{5GBeam} have developed a dataset for investigating beam selection techniques in the scenario of vehicle-to-infrastructure. The authors in \cite{MLPBeam} have made full use of the angle-of-arrival (AoA) information to design the multi-layer perceptron (MLP) to optimize the beam selection matrix.

Nevertheless, these black-box NNs generally have poor interpretability and generalization ability, and always require a great number of training samples. To address these issues, so-called deep-unfolding NNs have been developed in \cite{UnfoldSurvey,UnfoldTopic,AMP,Qiyu,Detection}, where the iterative optimization algorithms have been unfolded into a layer-wise structure that is similar to the NN. 
In addition, deep reinforcement learning (DRL) \cite{Nature} has also been widely applied to solve the high-dimensional non-convex optimization problems. Combining with the deep neural network (DNN), the deep Q-Network (DQN) achieves better performance than the conventional Q-learning. The DQN has been widely applied in mobile offloading \cite{DRLoff}, dynamic channel access \cite{DRLAccess}, radio access \cite{DRLradio}, resource allocation \cite{DRLresou2,DRLResou3}, and mobile-edge computing (MEC) \cite{DRLMEC}. In \cite{DRLResou3}, a multi-agent DRL approach has been developed to maximize the network utility in the heterogeneous cellular networks. The authors in \cite{DRLMEC} have developed a DRL-based online offloading framework for the MEC network.

To the best of our knowledge, the problem of joint beam selection and precoding design has not been fully investigated since it is an NP-hard problem with discrete variables and coupling constraints, which is difficult to find a satisfactory solution with low complexity. 
In this work, we propose a novel framework to jointly optimize the beam selection and digital precoding matrices.
Firstly, we formulate the beam selection problem as a Markov decision process (MDP) and a double deep Q-network (DDQN) algorithm is developed to solve it. We aim at selecting the optimal beams to maximize the sum-rate for served users and ensuring that the constraints of the beam selection matrix are satisfied. The base station (BS) is treated as an agent and the state, action, and reward function are carefully designed for the agent. Moreover, we reduce the dimension of state by exploiting the sparsity of beamspace channel and model the selection of each beam as an action.  

Given the beam selection matrix, the problem of the digital precoding design is a classic sum-rate maximization problem with power constraints that can be settled by the iterative WMMSE algorithm \cite{WMMSE}. It achieves satisfactory sum-rate performance but with high complexity since it requires the high-dimensional matrix inversion and a great number of iterations. To address this problem, a deep-unfolding NN is developed, where the iterative WMMSE algorithm is unfolded into a layer-wise structure. In this way, a much smaller number of iterations, i.e., the layers in the deep-unfolding NN, are applied to approach the iterative WMMSE algorithm, and the matrix inversion is avoided to reduce the computational complexity. In addition, some trainable parameters are involved to improve its sum-rate performance. 
Furthermore, a training method is proposed to jointly train the DRL-based NN and the deep-unfolding NN, which is totally different from the existing joint design scheme \cite{JointHybrid}. We also compare the computational complexity of the proposed joint NN design with benchmarks. Numerical results verify that our proposed jointly trained NN significantly outperforms the existing iterative algorithms with reduced computational complexity. In addition, the proposed jointly trained NN has much stronger robustness against the channel state information (CSI) errors. 
The contributions of this work can be summarized as follows.

\begin{itemize}
\item We formulate the problem of joint beam selection and precoding design for a downlink mmWave MU-MIMO system with DLA. To address this problem, we develop an efficient framework of joint NN design, where a DRL-based NN and a deep-unfolding NN are designed to jointly optimize the beam selection and digital precoding matrices.

\item As for the DRL-based NN, we formulate the beam selection problem as an MDP and a DDQN algorithm is developed to solve it. The state, action, and reward function are carefully designed for the agent BS, where we utilize the beamspace channel's characteristics and guarantee that the constraints of the beam selection matrix are satisfied. 

\item We develop a deep-unfolding NN with introduced trainable parameters to optimize the digital precoding matrix, where the iterative WMMSE algorithm is unfolded into a layer-wise structure. A much smaller number of iterations are applied to approximate the iterative WMMSE algorithm and the matrix inversion is avoided to reduce the computational complexity. 

\item A training method is proposed to jointly train the two NNs. Simulation results show that our proposed jointly trained NN significantly outperforms the existing iterative algorithms with much lower computational complexity. 
\end{itemize}

The rest of paper is structured as follows. Section \ref{SystemModel} introduces the system model of the downlink MU-MIMO system with DLA and formulates the problem mathematically. Section \ref{Framework} describes the framework of the joint NN design. The DRL-based NN designed for the beam selection matrix is presented in Section \ref{DRLbasedNN}. In Section \ref{DeepUnfolding}, the deep-unfolding NN is developed for the digital precoding matrix and the computational complexity is analyzed. We present the our experimental results in Section \ref{Simulation}. Finally, the paper is concluded in Section \ref{Conclusion}.
 
\emph{Notations:} Scalars, vectors, and matrices are respectively denoted by lower case, boldface lower case, and boldface upper case letters.
$\mathbf{I}$ represents an identity matrix and $\mathbf{0}$ denotes an all-zero matrix.
For a matrix $\mathbf{A}$, ${\bf{A}}^T$, $\mathbf{A}^*$, ${\bf{A}}^H$, and $\|\mathbf{A}\|$ denote its transpose, conjugate, conjugate transpose, and Frobenius norm, respectively.
For a vector $\mathbf{a}$, $\|\mathbf{a}\|$ represents its Euclidean norm.
$\mathbb{E}\{ \cdot \}$ denotes the statistical expectation.
$\Re\{ \cdot \}$ ($\Im\{ \cdot \}$) denotes the real (imaginary) part of a variable.
$\textrm{Tr}\{ \cdot \}$ denotes the trace operation.
$| \cdot |$ denotes the absolute value of a complex scalar and $\circ$ denotes the element-wise multiplication of two matrices, i.e., Hadmard product.
${\mathbb{C}^{m \times n}}\;({\mathbb{R}^{m \times n}})$ denotes the space of ${m \times n}$ complex (real) matrices.

\section{System Model and Problem Formulation}
\label{SystemModel}
In this section, we first introduce the system model for joint beam selection and precoding design and then formulate the optimization problem mathematically.

\subsection{System Model}

\begin{figure}[t]
\begin{centering}
\includegraphics[width=0.9\textwidth]{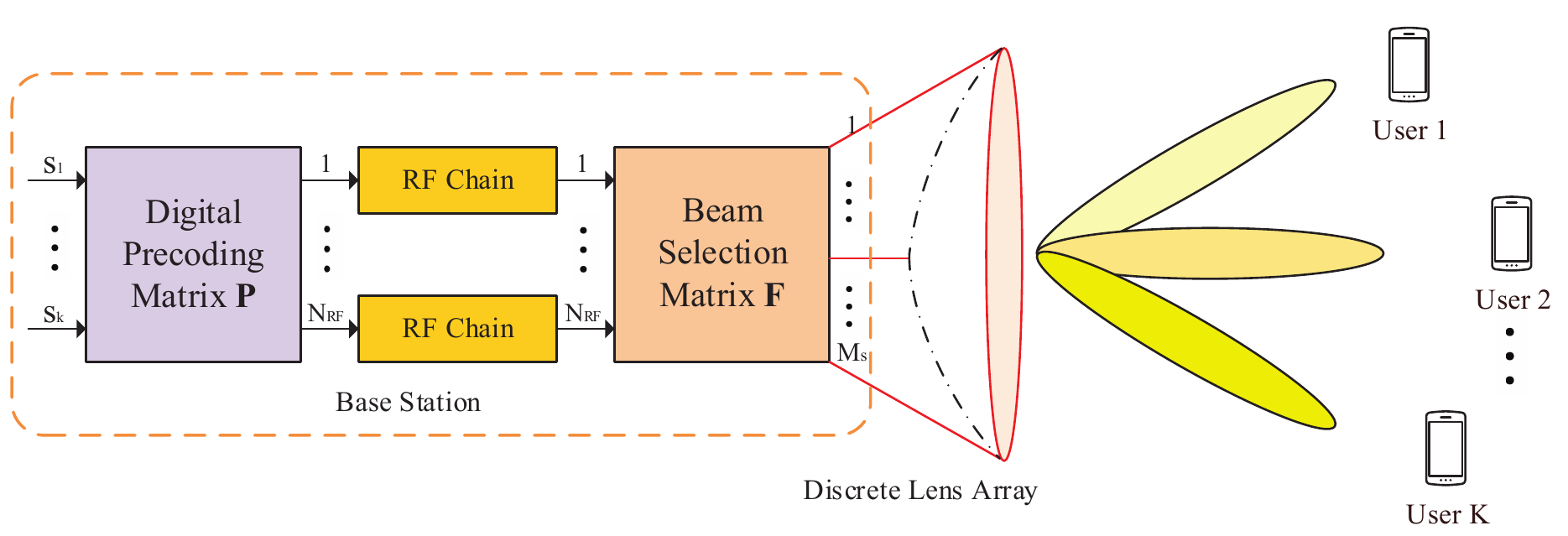}
\par\end{centering}
\caption{A downlink mmWave MU-MIMO system with discrete lens array.}
\label{System}
\end{figure}

As shown in Fig. \ref{System}, we study a downlink MU-MIMO system consisting of a BS equipped with a single-sided DLA, $M_s$ transmit antennas and $N_{RF} \ll M_{s}$ RF chains. The BS serves $K$ users simultaneously, each of which is equipped with a single antenna receiver. To ensure the spatial multiplexing gain for the $K$ users, the following relation should be satisfied: $N_{RF}\geq K$. The precoded data vector at the BS is expressed as
\begin{equation}
\mathbf{x}=\mathbf{P}\mathbf{s}=\sum\limits_{k=1}^{K}\mathbf{p}_k s_k,
\end{equation}
where $\mathbf{s}=[s_1, s_2, \cdots, s_K]^{T}$, $s_k$ denotes the transmit signal for user $k \in \mathcal{K}\triangleq \{1, 2, \cdots , K\}$ with zero mean and $ \mathbb{E}[\mathbf{s}\mathbf{s}^{H}]=\mathbf{I} $, and $\mathbf{P}=[\mathbf{p}_{1}, \mathbf{p}_{2}, \cdots, \mathbf{p}_{K}]\in \mathbb{C}^{N_{RF} \times K} $ denotes the precoding matrix and $\mathbf{p}_{k}$ is the digital precoding vector for user $k$. Then, the received signal vector $\mathbf{y}\in \mathbb{C}^{K \times 1}$ for all $K$ users is given by
\begin{equation}
\mathbf{y}=\mathbf{H}^{H}\mathbf{F}\mathbf{P}\mathbf{s}+\mathbf{n},
\end{equation}
where $\mathbf{H}\in \mathbb{C}^{M_{s}\times K} $ denotes the beamspace channel matrix, $\mathbf{F}\in \mathbb{C}^{M_{s}\times N_{RF}}$ is the beam selection matrix whose entries $f_{ij}, (i,j) \in \mathcal{T}$ are either $0$ or $1$, and $\mathcal{T}\triangleq \{ (i,j)|i=1, 2, \cdots, M_{s}, j=1, 2, \cdots, N_{RF} \}$. In addition, $\mathbf{n} \sim \mathcal{CN}(\mathbf{0}, \sigma^2 \mathbf{I}_{K})$ denotes the $K\times 1$ additive white Gaussian noise (AWGN) vector, where $\sigma^2$ represents the noise variance.

\subsection{Beamspace Channel Model}
The beamspace channel matrix $\mathbf{H}$ is modeled from the physical spatial MIMO channel
\begin{equation}
\mathbf{H}=[\mathbf{h}_1, \mathbf{h}_2, \cdots, \mathbf{h}_K]=[\mathbf{U}\mathbf{g}_1, \mathbf{U}\mathbf{g}_2, \cdots, \mathbf{U}\mathbf{g}_K],
\end{equation}
where $\mathbf{U}\in \mathbb{C}^{M_s \times M_s}$ is a discrete Fourier transformation (DFT) matrix and $\mathbf{g}_k \in \mathbb{C}^{M_s \times 1}$ denotes the spatial domain channel vector between the BS and user $k$. The DFT matrix $\mathbf{U}$ contains the array steering vectors of $M_s$ orthogonal beams \cite{RZhang}
\begin{equation}
\mathbf{U}=[\mathbf{a}(\varphi_1), \mathbf{a}(\varphi_2), \cdots, \mathbf{a}(\varphi_{M_s})]^H,
\end{equation}
where $\varphi_{m}=\dfrac{1}{M_s}(m-\dfrac{M_s+1}{2}), m = 1, 2, \cdots, M_{s}$, are the normalized spatial directions, $\mathbf{a}(\varphi_m)=\dfrac{1}{\sqrt{M_s}}[e^{-j2\pi \varphi_{m}i }]_{i \in \mathcal{I}} $ denotes the corresponding $M_s\times 1$ array steering vectors and $\mathcal{I}\triangleq \{n-\dfrac{M_{s}-1}{2} \big| n=0, 1, \cdots, M_{s}-1 \} $ is the index set of the array elements. Note that the columns of $\mathbf{U}$ are orthonormal, i.e., $\mathbf{U}^H\mathbf{U}=\mathbf{I}$.

Then, we apply the well-known channel model \cite{LowRF,NearOptim}
\begin{equation}
\mathbf{g}_k = \rho_k^{(0)}\mathbf{a}(\phi_k^{(0)}) + \sum\limits_{l=1}^L \rho_k^{(l)}\mathbf{a}(\phi_k^{(l)}), \label{channel}
\end{equation}
where $\rho_k^{(0)}\mathbf{a}(\phi_k^{(0)})$ and $\rho_k^{(l)}\mathbf{a}(\phi_k^{(l)})$ denote the line-of-sight (LoS) and the $l$-th non-line-of-sight (NLoS) channel vectors between the BS and user $k$, respectively. Moreover, $\rho_k^{(0)}$ and $\rho_k^{(l)}$ are the complex gains of the LoS and NLoS channels,  respectively, and $\phi_k^{(0)}$ and $\phi_k^{(l)}$ represent the spatial directions. For simplicity, we apply the 2D formulation that only considers the azimuth angel of departure (AoD). 
The number of NLoS components $L$ in \eqref{channel} is much less than $M_s$ since the number of scatters in a mmWave channel is limited, which leads to the sparse structure of $\mathbf{H}$.
In the channel model, $\mathbf{H}$ consists of $M_s$ beams and each row of $\mathbf{H}$ represents a beam vector.

\subsection{Problem Formulation}
We focus on the joint design of the beam selection matrix $\mathbf{F}$ and the digital precoding matrix $\mathbf{P}$ to maximize the downlink sum-rate of the system. The sum-rate maximization problem can be mathematically formulated as
\begin{subequations} \label{sum-rate max}
\begin{eqnarray}
& \max\limits_{ \{ \mathbf{F}, \mathbf{P} \} } &\sum\limits_{k=1}^{K}
\log \bigg(1 + \dfrac{| \mathbf{h}_k^{H}\mathbf{F}\mathbf{p}_{k} |^2}{\sum\limits_{i\neq k}^{K}| \mathbf{h}_k^{H}\mathbf{F}\mathbf{p}_{i} | + \sigma^2 } \bigg)  \label{sum-rate obj} \\
&\text{s.t.}  & \textrm{Tr}(\mathbf{P}^{H} \mathbf{F}^{T} \mathbf{F} \mathbf{P} )\leq P_{s}, \label{Power cons} \\
& & \sum\limits_{i=1}^{M_s} f_{ij}=1,  \forall j, \label{beam cons} \\
& & \sum\limits_{j=1}^{N_{RF}} f_{ij}\leq 1, \forall i, \label{RF cons} \\
& & f_{ij}\in \{0,1\}, \forall (i,j)\in \mathcal{T}, \label{discrete}
\end{eqnarray}
\end{subequations}
where $P_s$ is the transmit power budget at the BS. Note that the constraint \eqref{Power cons} can be simplified as $\textrm{Tr}(\mathbf{P}^{H}\mathbf{P})\leq P_{s}$ since $\mathbf{U}^{H} \mathbf{U} = \mathbf{I}_{M_{s}}$ and $\mathbf{F}^{T} \mathbf{F} = \mathbf{I}_{N_{RF}}$. The constraint \eqref{beam cons} guarantees that each RF chain feeds a single beam and the constraint \eqref{RF cons} ensures that each beam has been chosen for at most one RF chain. These constraints guarantee that we select $N_{RF}$ beams from $M_s$ beams to serve $K$ users. 
It is difficult to solve problem \eqref{sum-rate max} since it is an MINLP, which is non-convex and involves discrete variable $\mathbf{F}$. Therefore, we propose a joint NN design to solve it in the following sections.

\section{Framework of the Joint NN Design }
\label{Framework}
\begin{figure}[t]
\begin{centering}
\includegraphics[width=0.9\textwidth]{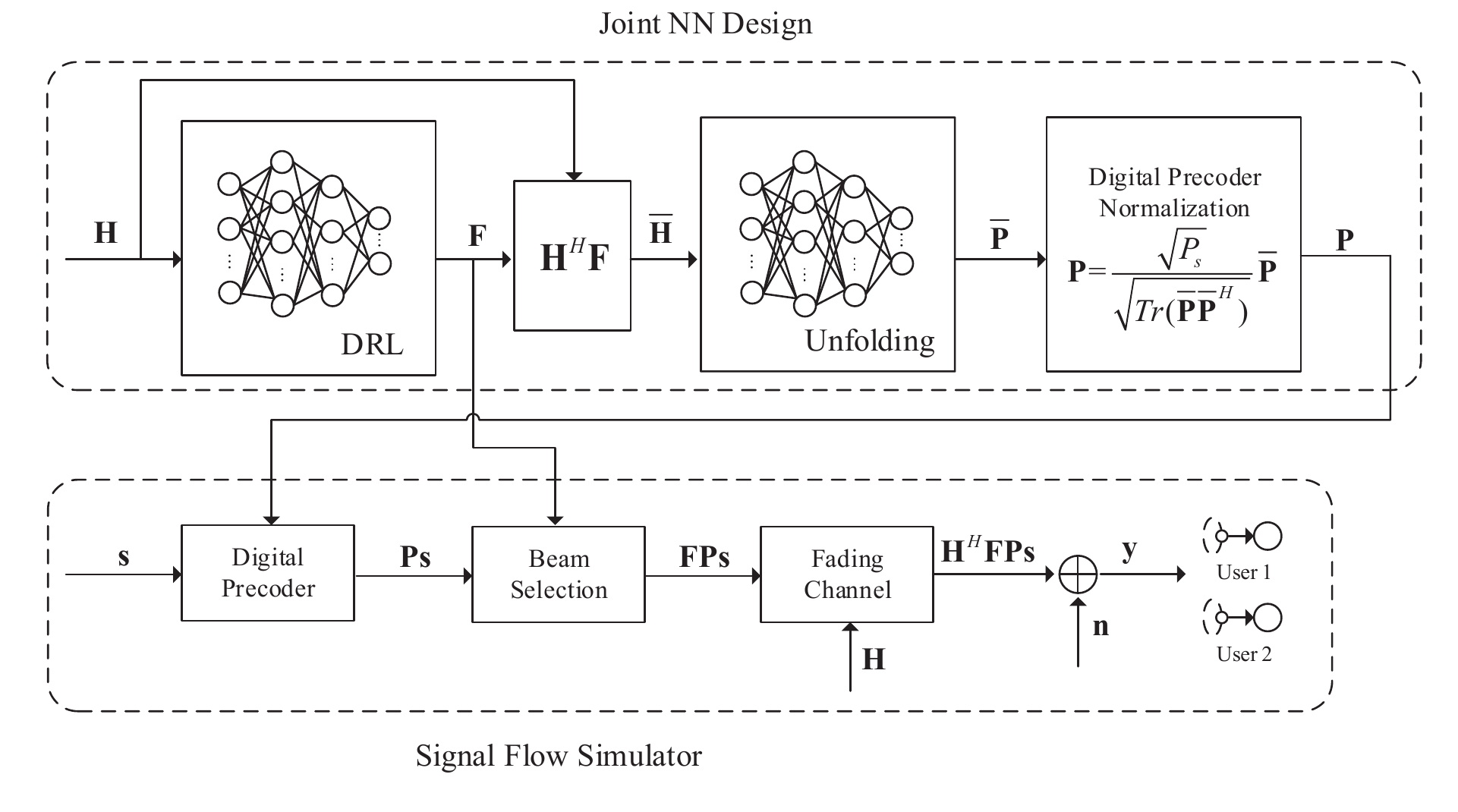}
\par\end{centering}
\caption{Proposed framework of joint beam selection and digital precoding design.}
\label{FrameNN}
\end{figure}

In this section, a framework of joint NN design for beam selection and digital precoding is developed. As shown in Fig. \ref{FrameNN}, the proposed framework includes two parts: joint NN design and signal flow simulator, which are elaborated as follows. 

\subsection{Joint NN Design}
To solve the MINLP \eqref{sum-rate max}, we propose a joint NN design consisting of a DRL-based NN for the beam selection matrix $\mathbf{F}$ and a deep-unfolding NN for the digital precoding matrix $\mathbf{P}$. Firstly, the channel matrix $\mathbf{H}$ is converted into a $2\times M_s \times K$ real-valued tensor, whose real part and imaginary part are stored separately. 
Then, we input $\mathbf{H}$ into the DRL-based NN with a novel DDQN architecture. It aims at selecting the optimal beams and maximizing the sum-rate for served users while ensuring that the constraints \eqref{beam cons}-\eqref{discrete} are satisfied. 
The DRL-based NN outputs the corresponding beam selection matrix $\mathbf{F}$. Then, the equivalent channel matrix $\bar{\mathbf{H}}^{H}=\mathbf{H}^{H}\mathbf{F} \in \mathbb{C}^{K\times N_{RF}}$ is obtained, which has much smaller dimension than $\mathbf{H}$. 

Given the equivalent channel matrix $\bar{\mathbf{H}}$, the problem regarding the digital precoding $\mathbf{\bar{P}}$ can be solved by the iterative WMMSE algorithm \cite{WMMSE} with significantly high complexity. To solve this problem, a deep-unfolding NN is developed, where the iterative WMMSE algorithm is unfolded into a layer-wise structure with some introduced trainable parameters. 
Specifically, $\bar{\mathbf{H}}$ is input into the deep-unfolding NN and the output is the unnormalized digital precoder $\mathbf{\bar{P}}$. 
Due to the structure of the investigated problem, we can see that the optimal solution makes that the power constraint \eqref{Power cons} always meet equality. To satisfy \eqref{Power cons}, we apply the following normalization layer to map $\mathbf{\bar{P}}$ into the digital precoder $\mathbf{P}$ as
\begin{equation}
\mathbf{P} = \dfrac{\sqrt{P_s}}{\sqrt{Tr(\mathbf{\bar{P}}\mathbf{\bar{P}}^{H})}}\mathbf{\bar{P}}.    \label{normalization}
\end{equation}
Finally, $\mathbf{H}$, $\mathbf{F}$, and $\mathbf{P}$ are substituted into the loss function
\begin{equation}
\mathcal{L}_{1}(\bm{\theta}) = -\dfrac{1}{N_s} \sum\limits_{n=1}^{N_s} f(\mathbf{H}_{n}, \mathbf{F}_{n}, \mathbf{P}_{n}; \bm{\theta} ),   \label{Loss Func}
\end{equation}
where $N_s$ denotes the size of training data set, $\bm{\theta}$ represents the trainable parameter of NN, and $f(\mathbf{H}_{n}, \mathbf{F}_{n}, \mathbf{P}_{n}; \bm{\theta} )$ indicates that the sum-rate \eqref{sum-rate obj} is achieved with the $n$-th channel realization.
In the training stage, with input $\mathbf{H}$, the DRL-based NN and the deep unfolding NN generate $\mathbf{F}$ and $\mathbf{P}$, respectively. We then perform the back propagation (BP) to jointly train the two NNs and employ the stochastic gradient descent (SGD) to update the trainable parameters. In addition, the outputs of the deep-unfolding NN will be substituted into \eqref{Loss Func}, which is part of the reward of the DRL-based NN and could help to train it efficiently. 

\subsection{Signal Flow Simulator}
In the prediction stage, the signal flow simulator imitates the process from the transmitting data symbol $\mathbf{s}$ to the detected symbol $\mathbf{y}$, over the wireless fading channel $\mathbf{H}$ with the AWGN $\mathbf{n}$ generated in the simulation environment. 
The channel $\mathbf{H}$ is input into the NNs and the outputs are $\mathbf{F}$ and $\mathbf{P}$. Thereafter, the digital precoder module and the beam selection module are replaced by the generated precoding matrix $\mathbf{P}$ and the beam selection matrix $\mathbf{F}$, respectively.

\section{DRL-based NN for Beam Selection}
\label{DRLbasedNN}
In this section, we present the DRL-based NN for beam selection after introducing the basic ideas of DRL.

\subsection{MDP and Value Function}
\subsubsection{MDP}
The MDP is defined by the quintuple $(\mathcal{S}, \mathcal{A}, \mathcal{R}, \mathcal{P}, \gamma)$, where $\mathcal{S}$ denotes the state space, $\mathcal{A}$ represents the action space, $\mathcal{R}: \mathcal{S} \rightarrow \mathbb{R}$ denotes the reward function, $\mathcal{P}: \mathcal{S} \times \mathcal{A} \times \mathcal{S} \rightarrow \mathbb{R}$ represents the state transition probability, and $\gamma \in (0,1)$ denotes the discount factor. 
Note that $s_{t}$, $a_{t}$, and $r_{t}$ denote the state, action, and reward at time step $t$, respectively.
Then, we have $a_{t}\sim \pi(\cdot|s_t)$, $s_{t+1}\sim p(\cdot|s_{t}, a_{t})$, $r_{t}\triangleq r(s_{t}, a_{t})\sim \mathcal{R}$, and the cumulative discounted reward $\sum_{t\geq 0}\gamma^{t}r_{t}$, where $p$ is the transition probability and $\pi$ denotes the policy, $\pi: \mathcal{S} \mapsto \mathcal{A}$. 

\subsubsection{Value Function}
The value function $V^{\pi}(s)$ at state $s$ is defined as the expected reward following the policy $\pi$
\begin{equation}
V^{\pi}(s) = \mathbb{E}\bigg\{ \sum\limits_{t\geq 0}\gamma^{t} r_{t} |s_{0} = s,\pi \bigg\},
\end{equation}
where the expectation $\mathbb{E}\{\cdot \}$ represents the empirical average over a batch of samples. Similarly, the action-value function $Q^{\pi}(s,a)$ denotes the expected reward obtained after taking action $a$ at state $s$ given the policy $\pi$, as
\begin{equation}
Q^{\pi}(s,a) = \mathbb{E}\bigg\{ \sum\limits_{t\geq 0}\gamma^{t} r_{t} |s_{0} = s, a_{0} = a, \pi \bigg\}.
\end{equation} 
The optimal policy $\pi^{\star}$ is obtained by the Bellman equation
\begin{equation}
Q^{\pi^{\star}}(s_t,a_t) = \max\limits_{\pi} \mathbb{E}\bigg\{ r_{t+1} + \gamma  \max\limits_{a} Q^{\pi}(s_{t+1},a)  |s_t, a_t \bigg\}.
\end{equation} 
The deep Q-learning is proposed to estimate the action-value function. Our objective in deep Q-learning aims at minimizing the loss of
\begin{equation} \label{DQNloss}
L(\bm{\theta}) = \mathbb{E}\bigg\{ \bigg( y_t - Q(s_{t+1},a;\bm{\theta}) \bigg)^{2} \bigg\},
\end{equation} 
where $\bm{\theta}$ denotes the parameter of the NN. We apply the DDQN \cite{DDQN} to learn the target value $y_t$ to avoid the overestimation and make the NN stable, which is expressed as 
\begin{equation}
y_t = \mathbb{E}\bigg\{ r_{t+1} + \gamma Q \bigg( s_{t+1}, \arg\max\limits_{a} Q(s_{t+1},a;\bm{\theta}) ;\bm{\theta'} \bigg) \bigg\},
\end{equation} 
where $\bm{\theta'}$ and $\bm{\theta}$ denote the parameters of the two NNs, which have been applied to select and evaluate the action, respectively. The gradient is expressed as  
\begin{equation} \label{DQN Grad estimator}
\nabla_{\bm{\theta}} L(\bm{\theta}) =  \mathbb{E}\bigg\{ r_{t+1} + \gamma Q \bigg( s_{t+1}, \arg\max\limits_{a} Q(s_{t+1},a;\bm{\theta});\bm{\theta'} \bigg) - Q(s_{t+1},a;\bm{\theta}) \nabla_{\bm{\theta}} Q(s_{t+1},a;\bm{\theta}) \bigg\}.
\end{equation} 

\subsection{DRL for Beam Selection}  
Firstly, we set up the process of beam selection as an MDP, where the BS is considered as the agent. Then, we carefully define the state, action, state transition, and reward function.
The difficulties and innovations of solving the MDP mainly include: (i) The beam selection problem is an NP-hard problem with discrete variables and coupling constraints, where the constraints \eqref{beam cons}-\eqref{discrete} must be satisfied. (ii) We aim at selecting the optimal beams and maximizing the sum-rate \eqref{sum-rate obj} by carefully designing the reward function. Hence, we need to make full use of the beamspace channel's characteristics and take into consideration the selected beam energy, the fairness of users, and the SINR. (iii) The state space is quite large due to the large number of $M_s$, which severely affects the convergence performance of NN. Thus, the dimension reduction is required to accelerate the convergence rate of NN without affecting the sum-rate performance. (iv) The DRL-based NN should be trained jointly with the deep-unfolding NN, which will be illustrated in detail in the next section.  

\subsubsection{MDP of Beam Selection}
The MDP is formulated mathematically as follows.
\begin{itemize}
\item Agent: The BS observes the current state $s_t$ and selects an action $a_t$ based on the policy $\pi$ to interact with the environment. The BS adjusts its policy $\pi$ based on the feedback reward from the environment. We aim at finding the optimal policy $\pi$ to maximize the expectation of the cumulative discounted reward $\mathbb{E}\big\{ \sum_{t\geq 0}\gamma^{t} r_{t} \big\}$.

\item State space: We convert the beamspace channel matrix $\mathbf{H}\in \mathbb{C}^{M_{s}\times K}$ into a $2\times M_s \times K$ real-valued tensor with the real part and imaginary part stored separately. The state space $\mathcal{S}\in \mathbb{C}^{3\times M_{s}\times K}$ consists of two parts: (i) the channel matrix $\mathbf{H}$ with dimension $2\times M_{s}\times K$; (ii) An indicator tensor with dimension $1\times M_{s}\times K$ to indicate whether a beam is selected. The elements of the indicator tensor are all initialized as $1$.

\item Action space: The action space is designed as the choice of the candidate beam, i.e., $\mathcal{A}=\{1, 2, \cdots, M_{s}\}$, and only one beam is selected at each time step $t$. The MDP has $N_{RF}$ time steps in an episode, i.e., $t\in \{1, 2, \cdots, N_{RF}\}$. By following this design, the constraints \eqref{beam cons} and \eqref{discrete} are satisfied.

\item State transition: Given the current state $s_{t}$ and the action $a_{t}=i$, the state transits to $s_{t+1}$ and the $i$-th row of the indicator tensor is set to be $\bm{0}$, while the channel matrix $\mathbf{H}$ keeps unchanged. 
\end{itemize}

\subsubsection{Reward Function} \label{subDRL}
Firstly, we need to ensure that the constraint \eqref{RF cons} is satisfied, i.e., a beam will be selected at most once. The agent will receive a penalty $\varrho$ if one beam is selected more than once, i.e., the selected beam whose corresponding locations in the indicator tensor are $\bm{0}$.
Then, if the selected beam has not been chosen before, the reward can contain the following parts, which evaluate the selected beam from different aspects. 
\begin{itemize}
\item First of all, the reward function could be defined as the $l_2$ norm of the selected beam vector at each time step since it shows the energy of the selected beam. In addition, $\mathbf{H}$ has the sparse structure and this part of the reward could avoid the case that the BS chooses the beam with low energy. 

\item Secondly, at the last time step $t=N_{RF}$, $N_{RF}$ beams are selected, thus we obtain the beam selection matrix $\mathbf{F}$. Since we aim at maximizing \eqref{sum-rate obj}, another part of the reward function at time step $t=N_{RF}$ is designed as
\begin{equation}
\sum_{k=1}^{K}
\log \bigg(1 + \dfrac{| \mathbf{h}_k^{H}\mathbf{F}\mathbf{p}_{k} |^2}{\sum_{i\neq k}^{K}| \mathbf{h}_k^{H}\mathbf{F}\mathbf{p}_{i} | + \sigma^2 } \bigg),  \label{reward} 
\end{equation}
where the digital precoding $\mathbf{p}_{k}$ is calculated by the deep-unfolding NN introduced in the next section.

\item Thirdly, since \eqref{reward} can only be applied at time step $t=N_{RF}$, we add the following expression as another part of the reward at each time step $t<N_{RF}$. It is an approximation of the SINR \cite{LowComp}, which measures the interference among the users,
\begin{equation}
\sum\limits_{k=1}^{K}\frac{ |h_{jk}^{t}|^2 }{ \sum\limits_{i\neq k}^{K} |h_{ji}^{t}|^2 + \sigma^2 }, \label{rewardSINR} 
\end{equation}
where $j$ is the index of the selected beam at time step $t$ and $h_{ji}^{t}$ represents the element in the $j$-th row and the $i$-th column of $\mathbf{H}$.

\item Furthermore, to avoid that there is no beam aligned for some users, whose sum-rate might approach zero, we add the average energy of each user as part of the reward at time step $t\geq \frac{N_{RF}}{2}$ to ensure the fairness among users
\begin{equation}
\sum\limits_{k=1}^{K} \frac{  \| \tilde{\mathbf{h}}_{k}^{t} \|^2 - \| \tilde{\mathbf{h}}_{k}^{t-1} \|^2 }{ \| \tilde{\mathbf{h}}^{t}_{k} \|^2 + \varepsilon } \big( \textit{sgn}( \delta - \| \tilde{\mathbf{h}}_{k}^{t-1} \|^2 ) +1 \big) . \label{rewardAve} 
\end{equation}
Note that $\tilde{\mathbf{H}}^{t}\in \mathbb{C}^{t\times K}$ denotes the channel matrix with $t$ selected beams consisting of all the beams chosen before the $t$-th time step, $\tilde{\mathbf{h}}_{k}^{t}\in \mathbb{C}^{t\times 1}$ represents the $k$-th column of $\tilde{\mathbf{H}}^{t}$, $\varepsilon$ is added here to avoid the numerical error, and $\delta$ is a given threshold. The term $ \| \tilde{\mathbf{h}}_{k}^{t} \|^2 - \| \tilde{\mathbf{h}}_{k}^{t-1} \|^2 $ represents the energy of the selected beams at time step $t$ for user $k$. The sign function $\textit{sgn}(\cdot)$ is added here since \eqref{rewardAve} is an extra reward for selecting a beam aligned for user $k$ when there is no beam allocated for it and the beam energy of user $k$ is very low, i.e., $\| \tilde{\mathbf{h}}_{k}^{t-1} \|^2 < \delta $. 
\end{itemize}

\subsubsection{Some Tricks for Improving the Performance}
\begin{figure}[t]
\begin{centering}
\includegraphics[width=0.4\textwidth]{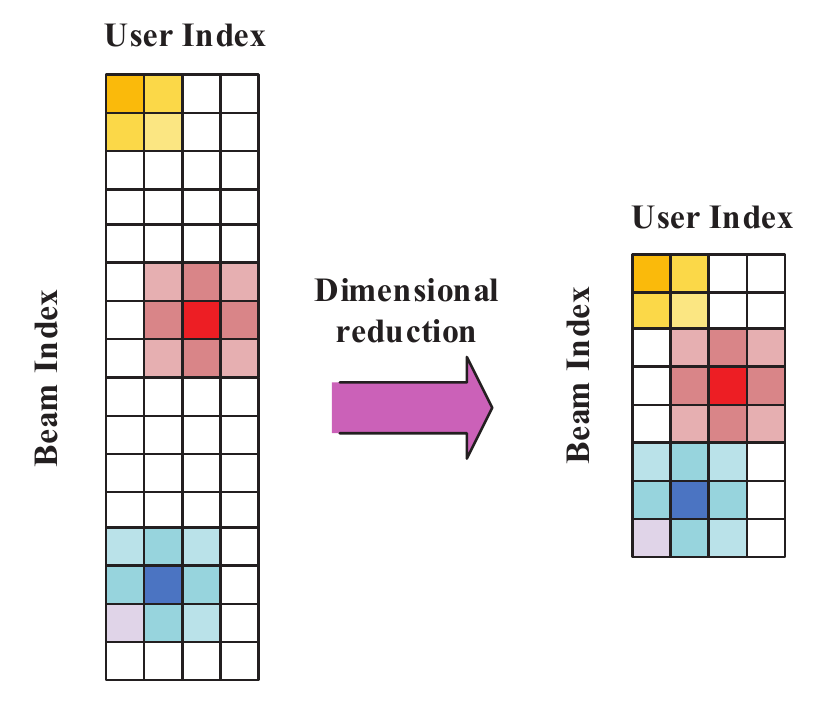}
\par\end{centering}
\caption{The dimension reduction of the beamspace channel, where the darker color represents the elements (beams) with larger magnitude (higher energy).}
\label{ReduceDim}
\end{figure}

\begin{figure}[t]
\begin{centering}
\includegraphics[width=0.99\textwidth]{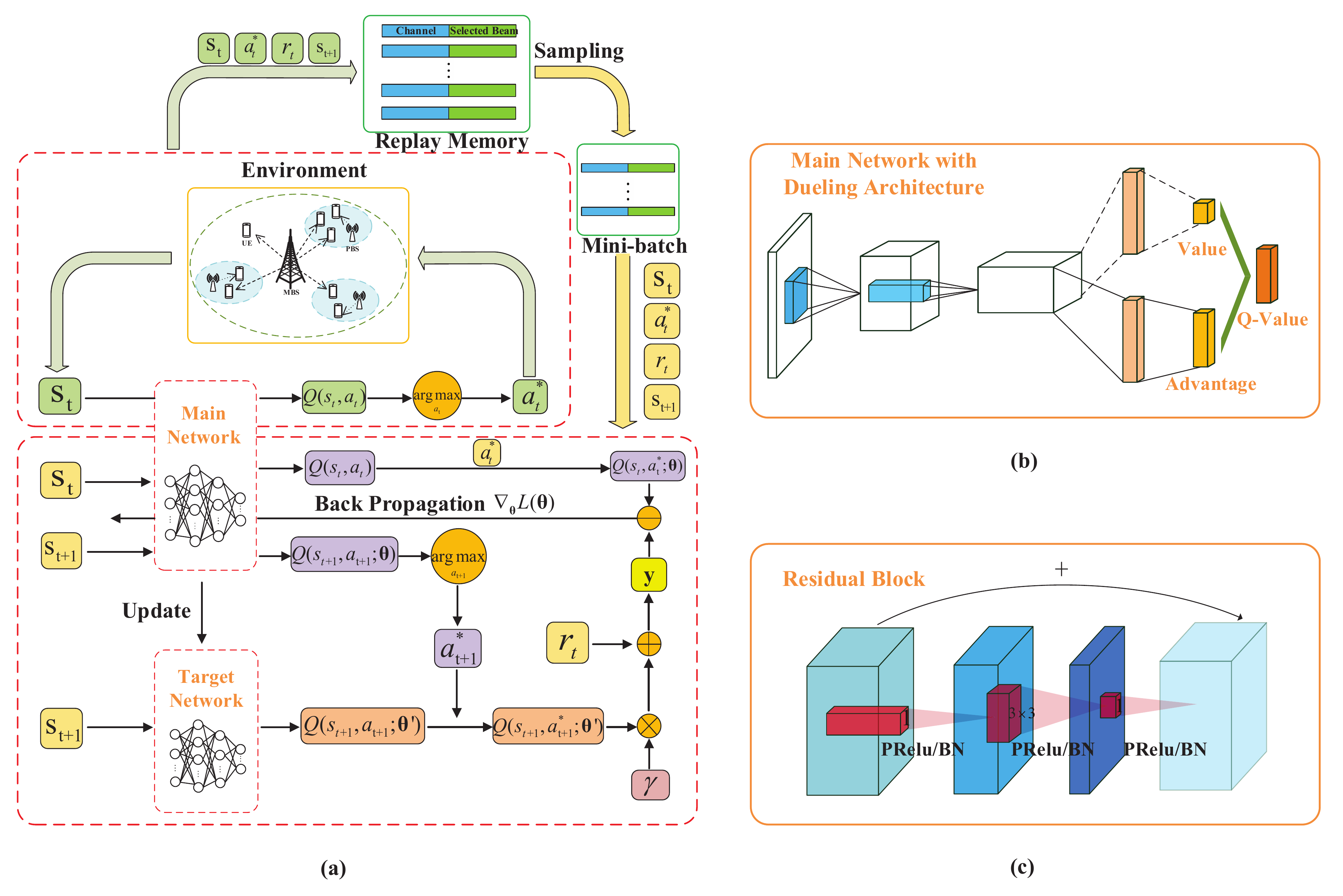}
\par\end{centering}
\caption{(a) Proposed DRL-based NN for beam selection; (b) Dueling Architecture; (c) Structure of the residual block in mobile net.}
\label{FrameDDQN}
\end{figure}

\begin{itemize}
\item \textit{Dimension reduction}: The dimension of beamspace channel matrix $\mathbf{H}\in \mathbb{C}^{M_{s}\times K}$ is very high due to the large number of $M_s$. We choose $\bar{M}_s (\bar{M}_s\ll M_s)$ highest energy beams to construct the beamspace channel matrix with reduced dimenson. Correspondingly, the dimensions of the state and action space are reduced to $3\times \bar{M}_{s}\times K$ and $\bar{M}_s$, respectively. 
The DRL-based NN is more stable and shows better convergence performance with lower dimensional state and action space.
The dimension reduction will not lead to performance degradation with a proper $\bar{M}_{s}$, due to the sparse structure of $\mathbf{H}$. 
The hyperparameter $\bar{M}_{s}$ needs to be fine tuned since a large number of $\bar{M}_{s}$ would lead to a high dimension while a small $\bar{M}_{s}$ might leave out some good beams.  
Fig. \ref{ReduceDim} presents the dimension reduction process, where the scenario of $K=N_{RF}=4$, $M_{s}=16$, and $\bar{M}_{s}=8$ is taken as an example. The beams with quite low energy are eliminated and the beamspace channel is reduced into $\bar{M}_s=8$ beams. 

\item \textit{Prioritized experience replay buffer \cite{DRLMEC}}: To overcome learning instability and reduce the correlation among training examples, we use a replay buffer $\mathcal{D}$ to store the transitions $(s_{t}, a_{t}, r_{t}, s_{t+1})$ obtained from the environment under the policy $\pi$ as shown in Fig. \ref{FrameDDQN}. A mini-batch of samples stored in $\mathcal{D}$ will be sampled for training based on the priority, where the samples with worse performance will be given higher priority. 

\item \textit{$\epsilon$-Greedy strategy \cite{DRLResou3}}: We apply the strategy that selects a random action $a_t$ with probability $\epsilon$, and selects the action $a_t = \max\limits_{a} Q^{\pi^{\star}}(s_t, a;\bm{\theta}) $ with probability $1-\epsilon$, where $\epsilon$ decays with time. It incorporates more candidate actions into the training samples.

\item \textit{Noisy network}: To enhance the exploration ability of the NN, we add the nosie to trainable parameters in the FC layers.
\end{itemize}

\subsection{Architecture of the DRL-based NN}

\subsubsection{DDQN}
Since the same NNs are applied to choose and evaluate the action in DQN, the Q-value function might be over-estimated. To avoid the overestimation, we apply the DDQN architecture \cite{DDQN} with two NNs to select and evaluate the action, which are referred to as the main network and target network, respectively. The procedure of the DRL-based NN with the DDQN architecture is presented in Fig. \ref{FrameDDQN} (a).
Particularly, the next state $s_{t+1}$ is employed by both the main network and target network to select the action and evaluate the value $Q(s_{t+1}, a_{t+1}; \bm{\theta}')$, respectively. Then, the target value $y$ is calculated with discount factor $\gamma$ and reward $r_t$. Finally, the error is computed by subtracting $y$ from the optimal value $Q(s_{t}, a_{t}^{*}; \bm{\theta})$, which is then backpropagated to update the trainable parameter $\bm{\theta}$.

\subsubsection{Dueling Architecture}
The Q-value function depicts how beneficial an action $a$ is taken the state $s$. We apply the dueling architecture to estimate the value function $V(s)$ and the advantage function $A(s, a)=Q(s, a)-V(s)$, where $A(s, a)$ describes the advantage of the action $a$ compared with the other actions. Thus, as shown in Fig. \ref{FrameDDQN} (b), the last layer of the DDQN is split into two subnetworks to evaluate $V(s)$ and $A(s, a)$ separately, and we restrict $\sum\limits_{a} A(s,a)=0$. Note that $V(s)$ can be interpreted as the mean value of the alternative actions at state $s$ and $A(s,a)$ evaluates the value of action $a$ compared to the mean value $V(s)$. Then, the action value function $Q(s, a)$ can be estimated by combining $V(s)$ with $A(s, a)$.

\subsubsection{Structure of the NN}
We adopt a well-known NN called ``Mobile Net" for the main network and target network, which consist of several cascaded residual blocks and an FC layer. Each residual block is comprised of the convolutional layers with $3\times 3$ filter and $1\times 1$ filter, the PRelu function, and the batch normalization (BN) layer, as presented in Fig. \ref{FrameDDQN} (c). Compared with the MLP, it has much smaller number of trainable parameters, which leads to faster convergence rate and more stable convergence performance.

Based on the aforementioned design, the training procedures of DRL-based NN for beam selection are described in \textbf{Algorithm \ref{DRLNN}}.

\begin{algorithm*}[t] \caption{Training procedures of DRL-based NN for beam selection} \label{DRLNN}
\begin{algorithmic}[1]
\begin{small} 
    \State Input: batch size $B$, learning rate $\eta$, replay period $Q$, the number of episode $J$, and the capacity of replay memory $D$;
   	\State Initialize the replay memory $\mathcal{D}=\emptyset$ and the action-value function $Q^{\pi}(s,a)$ with random parameter $\bm{\theta}$;  
	\For{episode = $1, 2, \cdots, J$}
	\State Observe the initial state $s_0$ and choose action $a_{0}\sim \pi_{\bm{\theta}}(s_{0})$;
	\For{time step t = $1, 2, \cdots, N_{RF}$}
	\State Selects a random action $a_t$ with probability $\epsilon$, and selects $a_t = \max\limits_{a} Q^{\pi^{\star}}(s_t, a;\bm{\theta}) $ with probability $1-\epsilon$;
	\State Execute the action $a_t$ and obtain the reward $r_t$ based on Section \ref{subDRL};
	\State Observe the next state $s_{t+1}$ and store transition $(s_{t}, a_{t}, r_{t}, s_{t+1})$ into the replay buffer $\mathcal{D}$;
	\EndFor
	\If{episode $=0$ mod $Q$}
	\State Sample a random mini-batch with size $B$ of transitions $(s_{t}, a_{t}, r_{t}, s_{t+1})$ from the replay buffer $\mathcal{D}$;	
	\State Compute loss function \eqref{DQNloss} and apply the SGD to update the trainable parameter $\bm{\theta}$ based on \eqref{DQN Grad estimator};
	\State Copy the trainable parameter $\bm{\theta}$ into the target network from time to time.
	\EndIf
	\EndFor 
\end{small}
\end{algorithmic}
\end{algorithm*}

\section{Deep-Unfolding NN for Precoding Design}
\label{DeepUnfolding}
In this section, we propose a deep-unfolding NN to design the digital precoder and analyze the computational complexity.

\subsection{Iterative WMMSE Algorithm}
The beam selection matrix $\mathbf{F}$ can be constructed based on the selected $N_{RF}$ beams and the equivalent channel matrix $\bar{\mathbf{H}}^{H}=\mathbf{H}^{H}\mathbf{F}$ is obtained.
Then, the sum-rate maximization problem \eqref{sum-rate max} is rewritten as follows for a given $\mathbf{F}$, 
\begin{subequations} \label{Prob fixF}
\begin{eqnarray}
& \max\limits_{ \{ \mathbf{p}_k \} } & \sum\limits_{k=1}^{K}
\log \bigg(1 + \dfrac{| \bar{\mathbf{h}}_k^{H} \mathbf{p}_{k} |^2}{\sum_{i\neq k}^{K}| \bar{\mathbf{h}}_k^{H} \mathbf{p}_{i} |^2 + \sigma^2 } \bigg) \label{sum-rate expre}  \\
&\text{s.t.}  & \sum\limits_{k=1}^{K} \textrm{Tr}(\mathbf{p}_k \mathbf{p}_k^{H} )\leq P_{s}, \label{Power const}
\end{eqnarray}
\end{subequations}
where $\bar{\mathbf{H}}\triangleq [ \bar{\mathbf{h}}_1, \bar{\mathbf{h}}_2, \cdots, \bar{\mathbf{h}}_K ]  \in \mathbb{C}^{N_{RF}\times K} $ and $\bar{\mathbf{h}}_k$ denotes the equivalent channel vector of user $k$, i.e.,  $\bar{\mathbf{h}}_k^{H}=\mathbf{h}_k^{H}\mathbf{F}$. It was demonstrated in \cite{WMMSE} that the MMSE problem \eqref{Prob WMMSE} shown below is equivalent to the sum-rate maximization problem \eqref{Prob fixF}, in the sense that the optimal solutions for these two problems are identical
\begin{subequations} \label{Prob WMMSE}
\begin{eqnarray}
& \min\limits_{ \{ \mathbf{p}_k, w_k, u_k \} } &\sum\limits_{k=1}^{K}
w_{k} e_{k} - \log w_{k}  \label{objWMMSE} \\
&\text{s.t.}  & \sum\limits_{k=1}^{K} \textrm{Tr}(\mathbf{p}_k \mathbf{p}_k^{H} )\leq P_{s},
\end{eqnarray}
\end{subequations}
where $u_k$ and $w_k$ are introduced auxiliary variables, and
\begin{equation}
e_k \triangleq |u_{k}\bar{\mathbf{h}}_{k}^{H}\mathbf{p}_{k}|^{2} - 2\Re (u_{k}\bar{\mathbf{h}}_{k}^{H}\mathbf{p}_{k}) + 1 + \sigma^2 |u_k|^{2} + \sum\limits_{i\neq k}|u_{k} \bar{\mathbf{h}}_{k}^{H}\mathbf{p}_{i}|^{2}. \notag
\end{equation}
The authors in \cite{WMMSE} have developed an iterative WMMSE algorithm to address this problem, where they alternatively optimize one variable of $\{u_{k}, w_{k}, \mathbf{p}_k\}$ with the other two fixed. The iterative closed-form expressions are given by
\begin{subequations} \label{WMMSE algo}
\begin{eqnarray}
&  & u_k = \big( \sum\limits_{i=1}^{K}\bar{\mathbf{h}}_{k}^{H}\mathbf{p}_{k}\mathbf{p}_{k}^{H}\bar{\mathbf{h}}_{k} + \sigma^{2} \big)^{-1} \mathbf{p}_{k}^{H}\bar{\mathbf{h}}_{k}, \quad \forall k, \label{iter u} \\
&  & w_k = ( 1-u_{k}\bar{\mathbf{h}}_{k}^{H}\mathbf{p}_{k} )^{-1}, \quad \forall k, \label{iter w}  \\
&  & \mathbf{p}_k = \bigg( \sum\limits_{k=1}^{K} w_{k}  \bar{\mathbf{h}}_{k} u_{k}^{*}u_{k}\bar{\mathbf{h}}_{k}^{H} + \lambda \mathbf{I}  \bigg)^{-1}\times \bigg( w_{k} \bar{\mathbf{h}}_{k} u_{k} \bigg), \quad \forall k, \label{iter p}
\end{eqnarray}
\end{subequations}
where $\lambda$ denotes the Lagrangian multiplier. The procedure of the iterative WMMSE algorithm is executing \eqref{iter u}, \eqref{iter w}, and \eqref{iter p} iteratively until \eqref{objWMMSE} converges. Based on the iterative WMMSE algorithm given by the expressions shown in \eqref{iter u}-\eqref{iter p}, we develop a novel deep-unfolding NN. 

\subsection{Deep-Unfolding NN for Precoding Design}
\begin{figure}[t]
	\begin{centering}
	\includegraphics[width=0.99\textwidth]{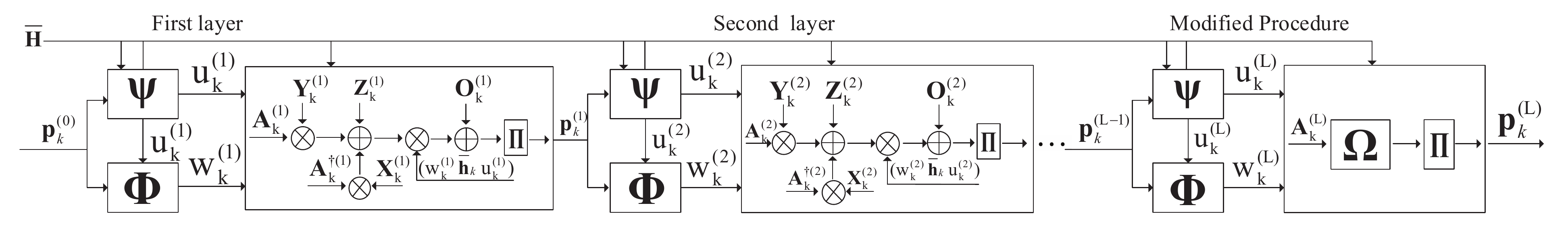}
	\par\end{centering}
	\caption{The architecture of deep-unfolding NN for digital precoding design.}
	\label{Unfolding}
\end{figure}

Then, we introduce the deep-unfolding NN, where the iterative WMMSE algorithm is unfolded into a layer-wise structure, as shown in Fig. \ref{Unfolding}. The computational complexity of the iterative WMMSE algorithm is high since there exists the matrix inversion operation in \eqref{iter p} and it usually requires a large number of iterations. We can apply a quite smaller number of iterations and avoid the matrix inversion to reduce the computational complexity. Moreover, some trainable parameters are involved to improve its sum-rate performance and avoid the bisection search of the Lagrangian multiplier.

Firstly, we define a novel non-linear function that takes the reciprocal of diagonal elements in matrix $\mathbf{A}$ and sets the off-diagonal elements to be $0$, which is denoted as $\mathbf{A}^{\nmid}$. We take a $3\times 3$ matrix as an example, 
\begin{equation}       
\mathbf{A} = \left[                
\begin{array}{ccc}  
a_{11} & a_{12} & a_{13}\\  
a_{21} & a_{22} & a_{23}\\
a_{31} & a_{32} & a_{33}\\
\end{array}
\right],   
\quad
\mathbf{A}^{\nmid} = \left[                 
\begin{array}{ccc}  
\frac{1}{a_{11}} & 0 & 0\\  
0 & \frac{1}{a_{22}}  & 0\\
0 & 0 & \frac{1}{a_{33}} \\
\end{array}
\right].                
\end{equation}
Note that $\mathbf{A}^{-1}=\mathbf{A}^{\nmid}$ when $\mathbf{A}$ is a diagonal matrix. We see that the diagonal elements of the matrices are much larger than the off-diagonal elements in the iterative WMMSE algorithm. Hence, $\mathbf{A}^{\nmid}$ is a good estimation of $\mathbf{A}^{-1}$. 
The matrix inversion $\mathbf{A}^{-1}$ is approximated by the combination of two parts: 
(i) $\mathbf{A}^{\nmid}\mathbf{X}$ with element-wise non-linear function $\mathbf{A}^{\nmid}$ and trainable parameter $\mathbf{X}$; 
(ii) By recalling the first-order Taylor expansion of $\mathbf{A}^{-1}$ at $\mathbf{A}_0$: $\mathbf{A}^{-1}=2\mathbf{A}_{0}^{-1}-\mathbf{A}_{0}^{-1}\mathbf{A}\mathbf{A}_{0}^{-1}$, we apply $\mathbf{A}\mathbf{Y}
+ \mathbf{Z}$ with trainable parameters $\mathbf{Y}$ and $\mathbf{Z}$. 

The architecture of the deep-unfolding NN is presented in Fig. \ref{Unfolding}, where $\bm{\Psi}$, $\bm{\Phi}$, and $\bm{\Omega}$ represent the iterative process of $u_k$ in \eqref{iter u}, $w_k$ in \eqref{iter w}, and $\mathbf{p}_k$ in \eqref{iter p}, respectively. We replace the matrix inversion $\mathbf{A}_k^{-1}$ as $\mathbf{A}_{k}^{\nmid} \mathbf{X}_k + \mathbf{A}_k \mathbf{Y}_k + \mathbf{Z}_k$, where $\mathbf{A}_k\triangleq \big( \sum\limits_{k=1}^{K} w_{k}  \bar{\mathbf{h}}_{k} u_{k}^{*}u_{k}\bar{\mathbf{h}}_{k}^{H} + \lambda_k \mathbf{I}  \big)$. 
To avoid the bisection search for the Lagrangian multiplier, we introduce $\lambda_k$ for each $\mathbf{p}_{k}$ as trainable parameters. Thus, in the deep-unfolding NN, the iterative process of $\mathbf{p}_k$ in \eqref{iter p} can be changed into
\begin{equation}
\mathbf{p}_{k}^{l} = \bigg( (\mathbf{A}_{k}^{l})^{\nmid}\mathbf{X}^{l}_{k} + \mathbf{A}_{k}^{l}\mathbf{Y}_{k}^{l} + \mathbf{Z}_{k}^{l} \bigg) w_{k}^{l} \bar{\mathbf{h}}_{k} u_{k}^{l} + \mathbf{O}_{k}^{l}, \quad \forall k, 
\end{equation}
where $(\cdot)^{(l)}$ denotes the $l$-th layer and $\{ \mathbf{X}_{k}^{l}, \mathbf{Y}_{k}^{l}, \mathbf{Z}_{k}^{l}, \mathbf{O}_{k}^{l}, \lambda_{k}^{l} \}$ are introduced trainable parameters. To satisfy the power constraint \eqref{Power const}, we apply the following projection operator
\begin{equation}
\bm{\prod}\{\mathbf{p}_k\} =
\begin{cases}
\mathbf{p}_k, & \sum\limits_{k=1}^{K} \textrm{Tr}(\mathbf{p}_k \mathbf{p}_k^{H} )\leq P_{s}, \\
\dfrac{\sqrt{P_s}}{\| \mathbf{P} \|}\mathbf{p}_k, & \textit{othewise},
\end{cases}  \label{projection}
\end{equation}
where $\mathbf{P}\triangleq [\mathbf{p}_{1}, \mathbf{p}_{2}, \dots, \mathbf{p}_{K}]$.
To improve the performance, we add a modified procedure in the last layer of the deep-unfolding NN, namely, employing the iterative process of $\mathbf{p}_k$ in \eqref{iter p}.

The training process of the deep-unfolding NN can be summarized as follows. Firstly, choose a batch of samples $\bar{\mathbf{H}}^{H}$ from the training dataset and input them into the deep-unfolding NN. Then, the digital precoding $\mathbf{P}$ is obtained and it is normalized to satisfy the power constraint. Next, $\{\bar{\mathbf{H}}^{H}, \mathbf{P}\}$ enters \eqref{sum-rate expre} to calculate the sum-rate, followed by computing the gradients of sum-rate \eqref{sum-rate expre} with respect to trainable parameters $\{ \mathbf{X}_{k}^{l}, \mathbf{Y}_{k}^{l}, \mathbf{Z}_{k}^{l}, \mathbf{O}_{k}^{l}, \lambda_{k}^{l} \}$. Finally, perform the back propagation and apply the SGD to update these parameters.  

In \textbf{Algorithm \ref{Process}}, we propose a training method to train the two NNs jointly, which is totally different from the existing joint design scheme \cite{JointHybrid} since the gradients cannot pass from the deep-unfolding NN to the DRL-based NN directly.
Compared with the conventional seperate training, the proposed joint training method connects the two NNs in two aspects: (i) the reward \eqref{reward} of the DRL-based NN is computed based on the digital precoding matrix $\mathbf{P}$ calculated by the deep-unfolding NN; (ii) the training samples $\bar{\mathbf{H}}^{H}$ of the deep-unfolding NN are generated by the DRL-based NN.  

\begin{algorithm*}[t]\caption{Joint training procedures of the two NNs} \label{Process}
\begin{algorithmic} [1]
\begin{small}
	\State Input the initialized parameters of the DRL-based NN in lines 1-2 of Algorithm 1. Input the parameters of the deep-unfolding NN, i.e., the number of layers $L$, the batch size $B$, and the tolerance of accuracy $\epsilon$ for convergence, and initialize its trainable parameters $\{ \mathbf{X}_{k}^{l}, \mathbf{Y}_{k}^{l}, \mathbf{Z}_{k}^{l}, \mathbf{O}_{k}^{l}, \lambda_{k}^{l} \}$ randomly;
	\While{the loss function of the DRL-based NN and the sum-rate do not converge}
	\For{episode = $1, 2, \cdots, J$}
	\State Execute lines 4-9 of Algorithm 1 for one episode to generate $(s_{t}, a_{t}, r_{t}, s_{t+1})$ and $\bar{\mathbf{H}}^{H}$;
	\State \parbox[t]{\dimexpr\linewidth-\algorithmicindent-\algorithmicindent}{Calculate the precoding matrix $\mathbf{P}$ via the forward propagation of the deep-unfolding NN with input $\bar{\mathbf{H}}^{H}$. Then, compute the reward \eqref{reward} based on $\bar{\mathbf{H}}^{H}$ and $\mathbf{P}$; } \vspace{0.1em}
	\State \parbox[t]{\dimexpr\linewidth-\algorithmicindent-\algorithmicindent}{Store $(s_{t}, a_{t}, r_{t}, s_{t+1})$ and $\bar{\mathbf{H}}^{H}$ into a buffer as the training dataset for the DRL-based NN and the deep-unfolding NN, respectively.}  \vspace{0.1em}
	\EndFor
	\If{episode $=0$ mod $Q$}
	\State \textbf{Train DRL-based NN}: Execute lines 11-13 of Algorithm 1;  
	\State \parbox[t]{\dimexpr\linewidth-\algorithmicindent-\algorithmicindent}{ \textbf{Train deep-unfolding NN}: Choose a batch of samples $\bar{\mathbf{H}}^{H}$ from the training dataset and input them into the deep-unfolding NN to perform the forward propagation. Compute the gradients of sum-rate \eqref{sum-rate expre} with respect to trainable parameters and perform the back propagation with the SGD to update these parameters. } \vspace{0.1em}
	\EndIf
	\EndWhile
\end{small}
\end{algorithmic}
\end{algorithm*}

\subsection{Computational Complexity}
Firstly, we analyze the computational complexity of the proposed joint NN design, which consists of two parts: the DRL-based NN and the deep-unfolding NN. The complexity of the DRL-based NN is $\mathcal{O} \big(\sum\limits_{l=1}^{L-1} Q_{l}^{2}S_{l}^{2}C_{l-1}C_{l} + KN_{RF}C_{L-1}C_{out} \big)$, where $L$ is the number of layers, $S_{l}$ represents the size of convolutional kernel, $C_{l}$ is the number of channels in the $l$-th layer, $C_{out}$ denotes the output size of the FC layer, and $Q_{l}$ denotes the output size of the $l$-th layer, which depends on the input size, padding number, and stride. The complexity of the deep-unfolding NN is $\mathcal{O} \big(I_{n} (K^{2}N_{RF}^{2}+KN_{RF}^{2.37}) \big)$, where $I_{n}$ is the number of layers. Compared to the complexity of the iterative WMMSE algorithm $\mathcal{O}(I_{m}(K^{2}N_{RF}^{2}+KN_{RF}^{3}))$, the deep-unfolding NN has much lower complexity for two reasons: (i) The number of layers in the deep-unfolding NN is smaller than that of the iterative WMMSE algorithm, i.e., $I_{n}<I_{m}$; (ii) The iterative WMMSE algorithm involves the matrix inversion with complexity $\mathcal{O}(N_{RF}^{3})$, while the deep-unfolding NN simply requires matrix multiplication with complexity $\mathcal{O}(N_{RF}^{2.37})$. 
In Table \ref{Complexity}, we present the computational complexity of the following schemes:
\begin{itemize}
\item Joint NN design: The proposed joint design of DRL-based NN and deep-unfolding NN;
\item PDD: The PDD-based iterative algorithm developed in \cite{RGuo};
\item MM-WMMSE: The maximum magnitude beam selection method with the iterative WMMSE precoding algorithm;
\item IA-ZF: The interference-aware beam selection method with the ZF precoding algorithm developed in \cite{NearOptim}; 
\item MS-ZF: The maximization of SINR selection method with ZF precoding proposed in \cite{LowRF};
\item FD-ZF: The fully digital ZF precoding; 
\item FD-WMMSE: The fully digital WMMSE precoding \cite{WMMSE}. 
\end{itemize}

In Table \ref{Complexity}, $I_{p_{1}}$ and $I_{p_{2}}$ represent the number of inner and outer iterations of the PDD-based algorithm, and $I_{m}$ and $I_{w}$ are the number of iterations of the MM-WMMSE and FD-WMMSE, respectively. Based on the complexity of beam selection and precoding, the overall complexity in Table \ref{Complexity} is calculated under the condition $M_{s}\gg N_{RF}\geq K$.
Note that the overall complexity of the proposed joint NN design is unrelated to $M_s$, which reduces the complexity to some extent. 
From Table \ref{Complexity}, we can see that the FD-WMMSE precoding has very high computational complexity since it requires the same number of RF chains as transmit antennas, which also results in excessive hardware costs. In contrast, the proposed joint NN design achieves lower complexity but still maintains performance levels that approach the FD precoding, which will be presented in the simulation results. Moreover, the IA-ZF and MS-ZF algorithms have a generally lower computational complexity than the PDD and MM-WMMSE algorithms, but their performance is not as good. Therefore, our proposed joint NN design achieves a practical trade-off between complexity and performance.

\begin{table*}[t]
\caption{Computational complexity of the analyzed schemes.}
\begin{center}
\centering
\begin{tabular}{|c|c|c|c|}
\hline
\textbf{Algorithms}    & \textbf{Beam selection} & \textbf{Precoding} & \textbf{Overall} ($M_{s}\gg N_{RF}\geq K$)  \\
\hline
\textbf{Joint NN design}  & \tabincell{c}{$\mathcal{O} \big(\sum\limits_{l=1}^{L-1} Q_{l}^{2}S_{l}^{2}C_{l-1}C_{l}$ \\ $+ KN_{RF}C_{L-1}C_{out} \big)$} & $\mathcal{O} \big(I_{n}(K^{2}N_{RF}^{2}+KN_{RF}^{2.37}) \big)$ & \tabincell{c}{$\mathcal{O} \big( \sum\limits_{l=1}^{L-1} Q_{l}^{2}S_{l}^{2}C_{l-1}C_{l}$ \\ $+I_{n}(K^{2}N_{RF}^{2}+KN_{RF}^{2.37}) \big)$}  \\
\hline
\textbf{PDD}  & $\mathcal{O} ( I_{p_{1}}I_{p_{2}}M_{s}^{2}N_{RF}K ) $  &$\mathcal{O}(I_{p_{1}}I_{p_{2}}M_{s}^{3} )$  & $\mathcal{O} \big( I_{p_{1}}I_{p_{2}}(M_{s}^{2}N_{RF}K + M_{s}^{3} ) \big)$  \\
\hline
\textbf{MM-WMMSE}  & $\mathcal{O}(M_{s}\log M_{s})$ & $\mathcal{O}(I_{m}(K^{2}N_{RF}^{2}+KN_{RF}^{3}))$ & $\mathcal{O}(M_{s}\log M_{s}+I_{m}KN_{RF}^{3})$  \\
\hline
\textbf{IA-ZF}  & $\mathcal{O}(N_{RF}M_{s})$ & $\mathcal{O}(N_{RF}K^{2})$ & $\mathcal{O}(N_{RF}M_{s}+N_{RF}K^{2})$ \\
\hline
\textbf{MS-ZF}  & $\mathcal{O}(N_{RF}KM_{s}^{2})$ &  $\mathcal{O}(N_{RF}K^{2})$ & $\mathcal{O}(N_{RF}KM_{s}^{2})$ \\
\hline
\textbf{FD-ZF}  & $-$ & $\mathcal{O}(M_{s}K^{2})$ & $\mathcal{O}(M_{s}K^{2})$ \\
\hline
\textbf{FD-WMMSE}  & $-$ & $\mathcal{O}(I_{w}KM_{s}^{3})$ & $\mathcal{O}(I_{w}KM_{s}^{3})$ \\
\hline
\end{tabular}
\end{center}
\label{Complexity}
\end{table*}

\section{Simulation Results}
\label{Simulation}

In this section, we evaluate the performance of the aforementioned algorithms by performing numerical simulations. 

\subsection{Simulation Setup}
The sum-rate performance of the proposed joint NN design is evaluated in the testing dataset after the NN converges. The system configuration is described as follows. Unless otherwise specified, the BS is equipped with a DLA with $M_{s}=256$ antennas and $N_{RF}=20$ RF chains to serve $K=20$ users. We set the transmission power and the noise to be $30$ dBm and $-10$ dBm, respectively. The parameters of the channel model \eqref{channel} are selected based on \cite{GCBeam}: (i) We adopt one LoS link and $L=3$ NLoS links; (ii) $\rho_k^{(0)}\sim \mathcal{CN}(0,1)$ and $\rho_k^{(l)}\sim \mathcal{CN}(0,0.1)$ for $l=1,2,3$; (iii) $\phi_k^{(0)}$ and $\phi_k^{(l)}$ are generated randomly within $[-1, 1]$; (iv) $\rho_k^{(0)}$, $\rho_k^{(l)}$, $\phi_k^{(0)}$, and $\phi_k^{(l)}$ are statistically independent.
The parameters for the DRL-based NN are set as: discount rate $\gamma=0.9$, buffer size $D=16000$, batch size $B=40$, replay period $Q=10$, and the penalty of repeated selected beam $\varrho = -50$. The number of layers for the deep-unfolding NN is set to be $L=6$. The number of training and testing samples are $5000$ and $1000$, respectively. 
We run the simulation on the platform of ``Pytorch 1.5.0" with ``Python 3.6". Since the platform cannot handle complex matrices, we convert $\mathbf{A}\in \mathbb{C}^{a\times b}$ into a $2\times a \times b$ real-valued tensor with the real part $\Re\{\mathbf{A}\}$ and imaginary part $\Im\{\mathbf{A}\}$ stored separately. Then, the real part and imaginary part of the multiplication of two complex matrices $\mathbf{A}$ and $\mathbf{B}$ can be written as
$(\Re\{\mathbf{A}\} \Re\{\mathbf{B}\}-\Im\{\mathbf{A}\} \Im\{\mathbf{B}\})$ and $(\Re\{\mathbf{A}\} \Im\{\mathbf{B}\}+\Im\{\mathbf{A}\} \Re\{\mathbf{B}\})$, respectively.
For the inversion and determinant operations of complex matrix, we override the automatic differential of ``Pytorch" by calculating the closed-form gradients employing the platform ``Numpy" based on the formulas
\begin{equation}
d\log \det(\mathbf{X})=\textrm{Tr}(\mathbf{X}^{-1}d\mathbf{X}), \quad  d\textrm{Tr}(\mathbf{X}^{-1})=-\textrm{Tr}(\mathbf{X}^{-1} (d\mathbf{X})\mathbf{X}^{-1}),
\end{equation}
which provide much higher accuracy. 

\subsection{Convergence Performance of the Joint NN Design}

\begin{figure}[!t]
\centering
\subfloat[]{\centering \scalebox{0.55}{\includegraphics{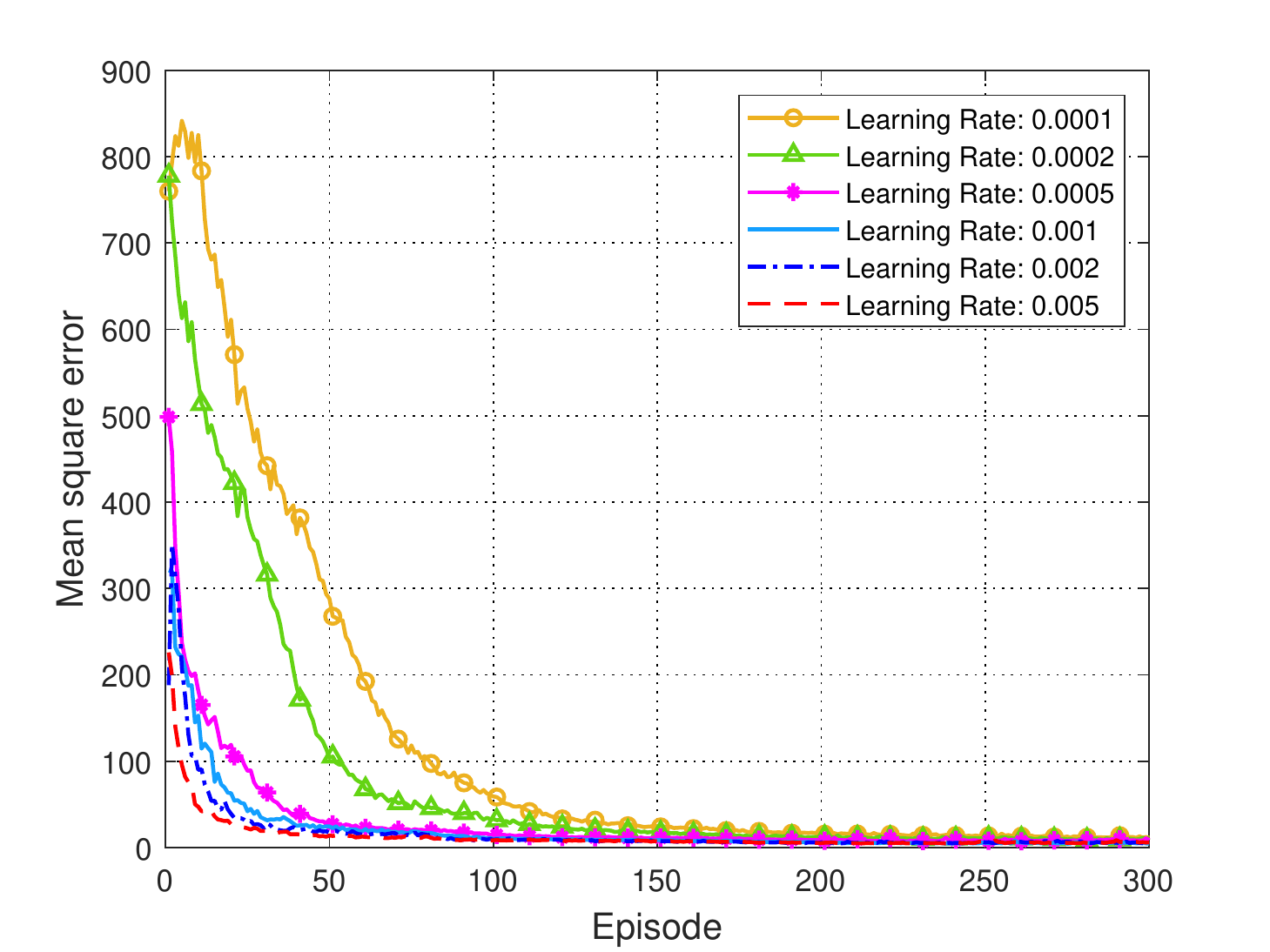}} }
\subfloat[]{\centering \scalebox{0.55}{\includegraphics{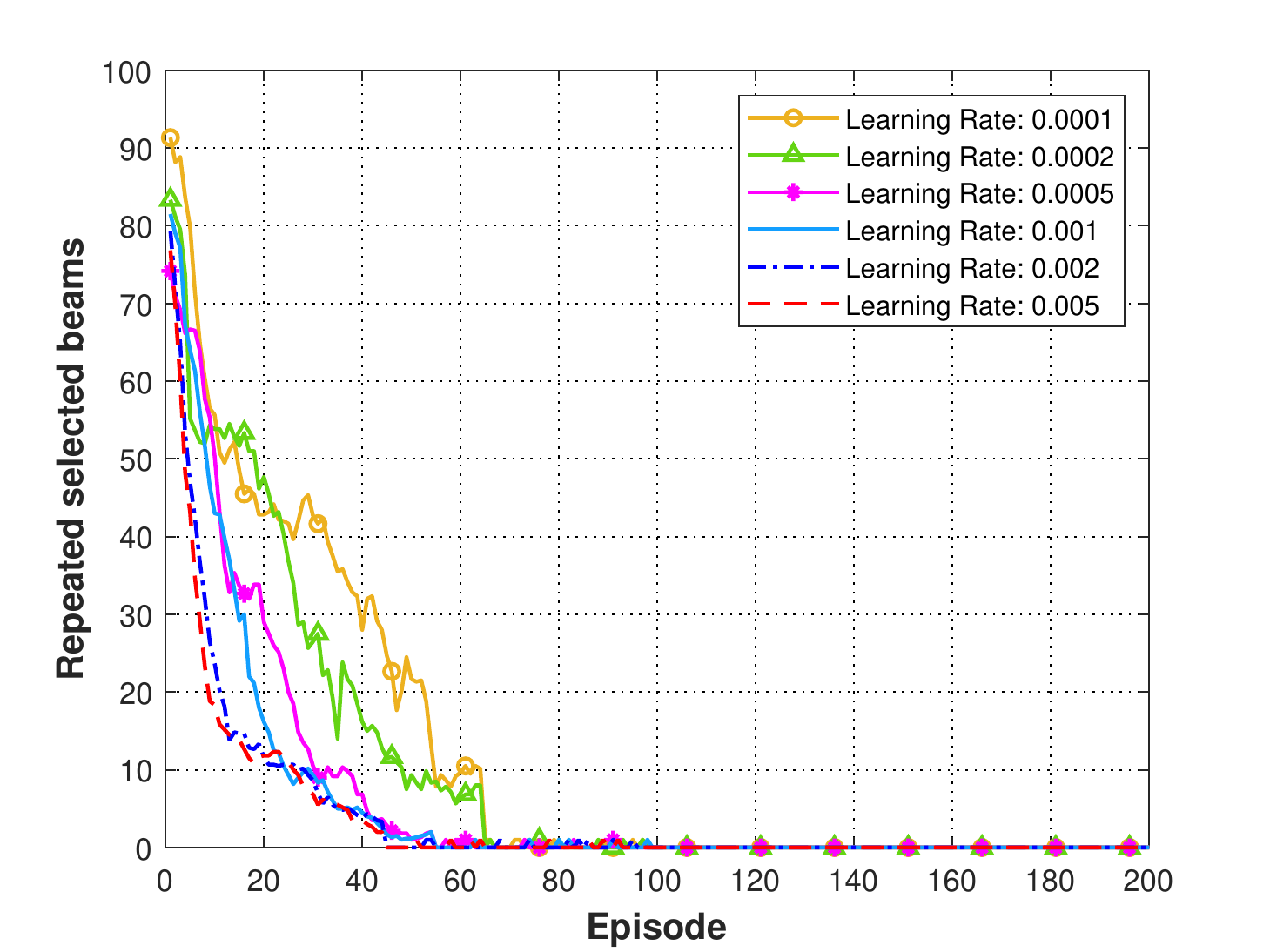}}}	
\caption{(a) Convergence performance of the loss function (MSE) with different learning rates; (b) The constraint violation (number of repeated selected beams).}
\label{LossBeam}
\end{figure}

We first present the convergence performance of the proposed joint NN design. Fig. \ref{LossBeam}(a) shows the convergence performance of the loss function, i.e., mean square error (MSE), with different learning rates. It is observed that a smaller learning rate achieves better MSE performance, while a larger learning rate results in a faster convergence speed. Fig. \ref{LossBeam}(b) presents the constraint violation, i.e., the number of repeated selected beams for $50$ samples with $1000$ selected beams. The constraint \eqref{RF cons} is satisfied when it equals $0$ and a larger number means that it violates the constraint \eqref{RF cons} more severely. We can see that the constraint is severely violated at the beginning and is well satisfied after around $100$ episodes, where a larger learning rate results in a faster convergence speed.

\begin{figure}[t]
\begin{centering}
\includegraphics[width=0.55\textwidth]{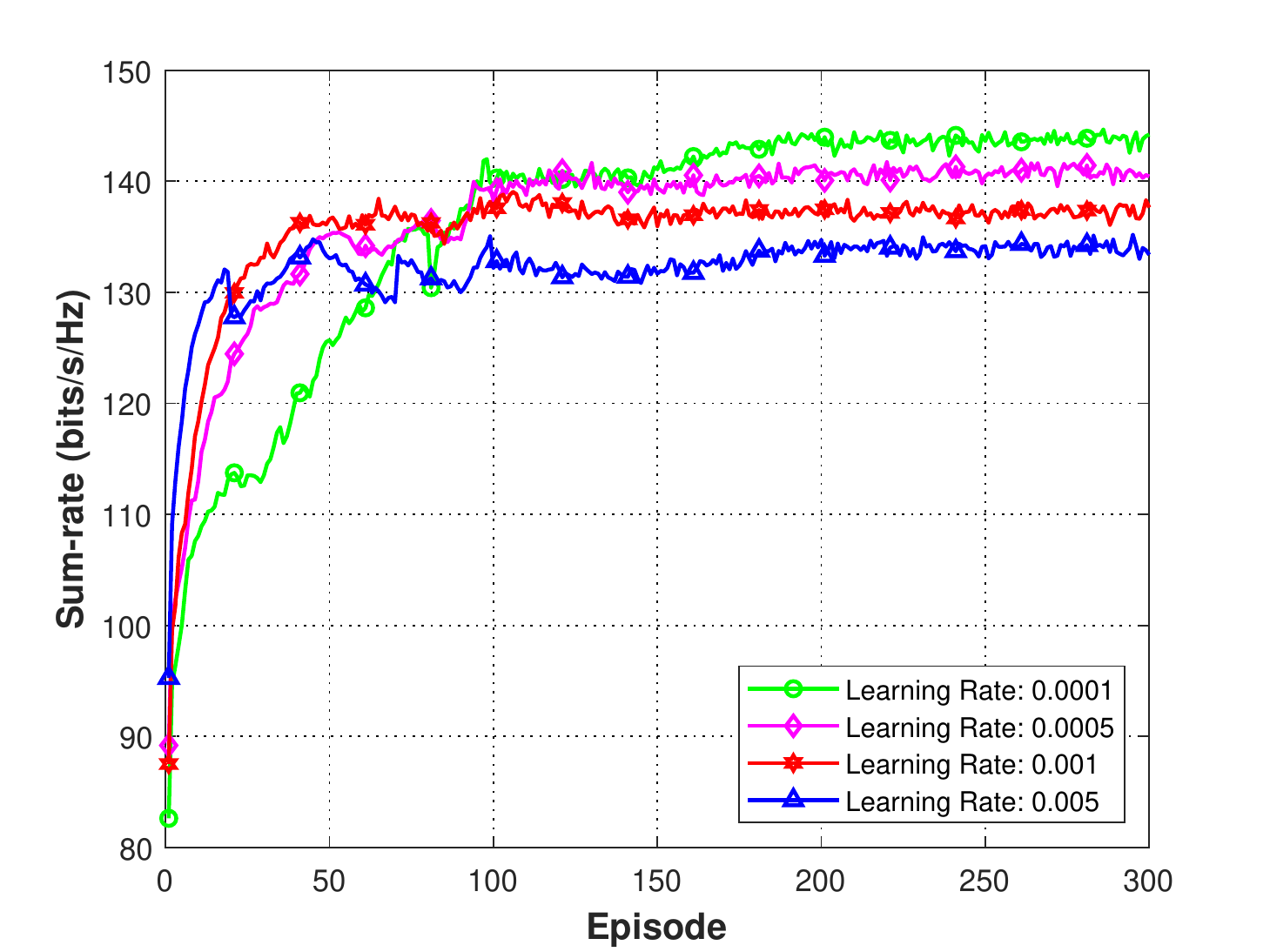}
\par\end{centering}
\caption{Convergence performance of the sum-rate with different learning rates.}
\label{SumRate}
\end{figure}

Fig. \ref{SumRate} presents the sum-rate performance of the proposed joint NN design achieved by different learning rates. Similar to the MSE in Fig. \ref{LossBeam}(a), a smaller learning rate achieves better sum-rate performance with more stable convergence performance, and a larger learning rate leads to a faster convergence speed.

\subsection{Sum-Rate Performance}
\begin{figure}[t]
\begin{centering}
\includegraphics[width=0.55\textwidth]{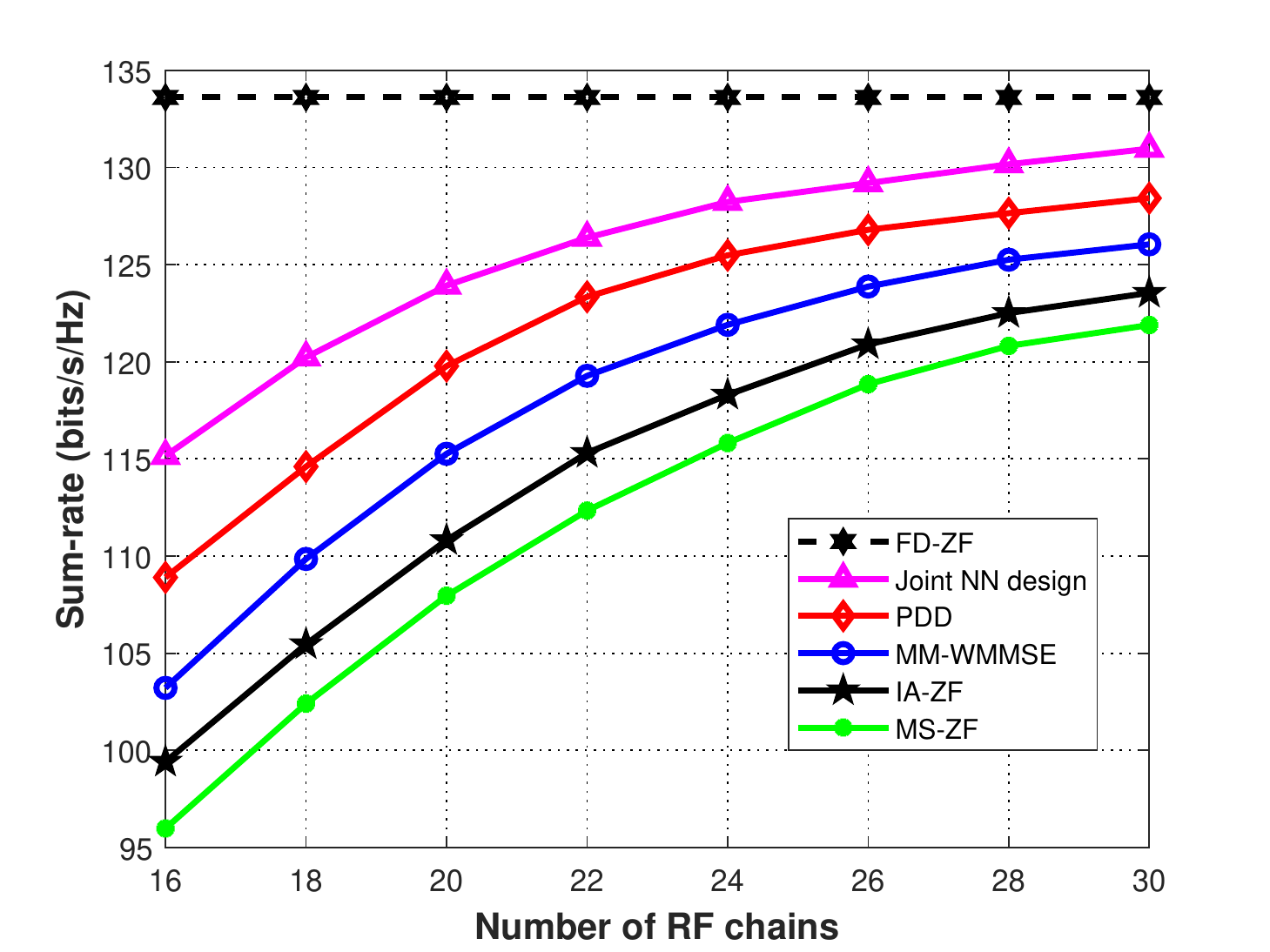}
\par\end{centering}
\caption{Achievable system sum-rate versus the number of RF chains $N_{RF}$. }
\label{RNrf}
\end{figure}

Fig. \ref{RNrf} compares the sum-rate versus the number of RF chains achieved by the schemes mentioned above where $K=16$. The sum-rate performance of all schemes increases monotonically with the number of RF chains. It is obvious that the gap between the proposed joint NN design and the FD-ZF precoding reduces with the increase of $N_{RF}$. Moreover, the joint NN design can achieve the sum-rate performance approaching the FD-ZF precoding with a small number of RF chains, i.e., $N_{RF}=30$. When $N_{RF}$ becomes larger, the sum-rate reaches its bottleneck and increases more slowly. From the results, the best performance is achieved by the FD-ZF precoding, followed by the joint NN design, PDD, MM-WMMSE, IA-ZF, and MS-ZF schemes. The PDD and MM-WMMSE outperform the IA-ZF and MS-ZF mainly because the former two schemes apply the iterative WMMSE algorithm for digital precoding, which achieves better performance than the ZF precoding. In addition, the PDD outperforms the MM-WMMSE and the IA-ZF outperforms the MS-ZF since the PDD and IA select better beams than the MM and MS.   

\begin{figure}[t]
\begin{centering}
\includegraphics[width=0.55\textwidth]{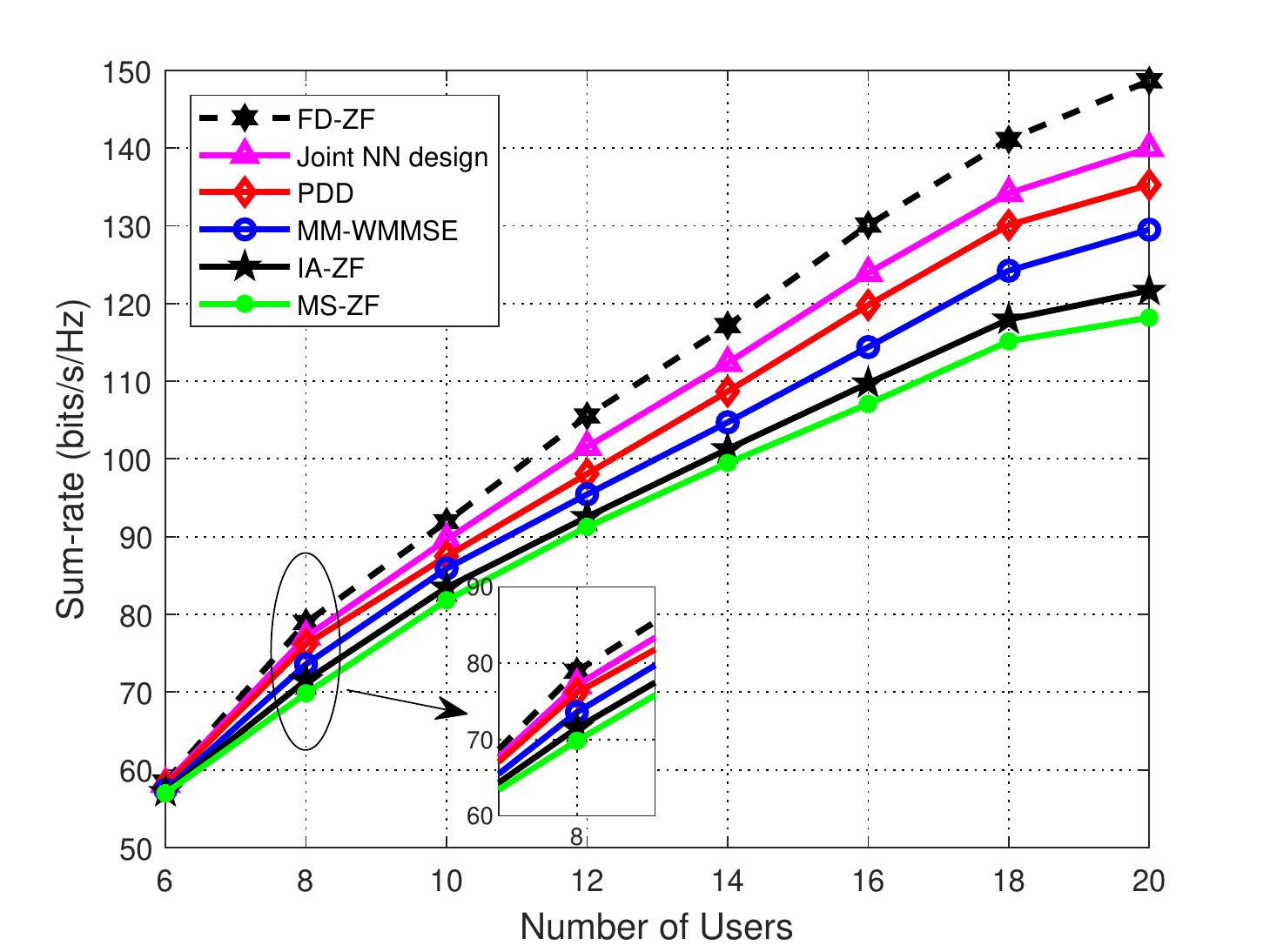}
\par\end{centering}
\caption{Achievable system sum-rate versus the number of users $K$.}
\label{RK}
\end{figure}

Fig. \ref{RK} presents the achievable sum-rate versus the number of users where $N_{RF}=20$. We see that the sum-rate performance achieved by all these algorithms increases with $K$. The joint NN design achieves the performance close to the FD-ZF precoding, where the gap increases with $K$. From the results, the proposed joint NN design delivers better sum-rate performance than the PDD, MM-WMMSE, IA-ZF, and MS-ZF schemes, which demonstrates its superiority to mitigate the multi-user interference. Besides, the performance gap between the proposed joint NN design and the other schemes escalates upon the increasing number of users, which shows that the joint NN design has potential for applications with a large number of users.

\begin{figure}[t]
\begin{centering}
\includegraphics[width=0.55\textwidth]{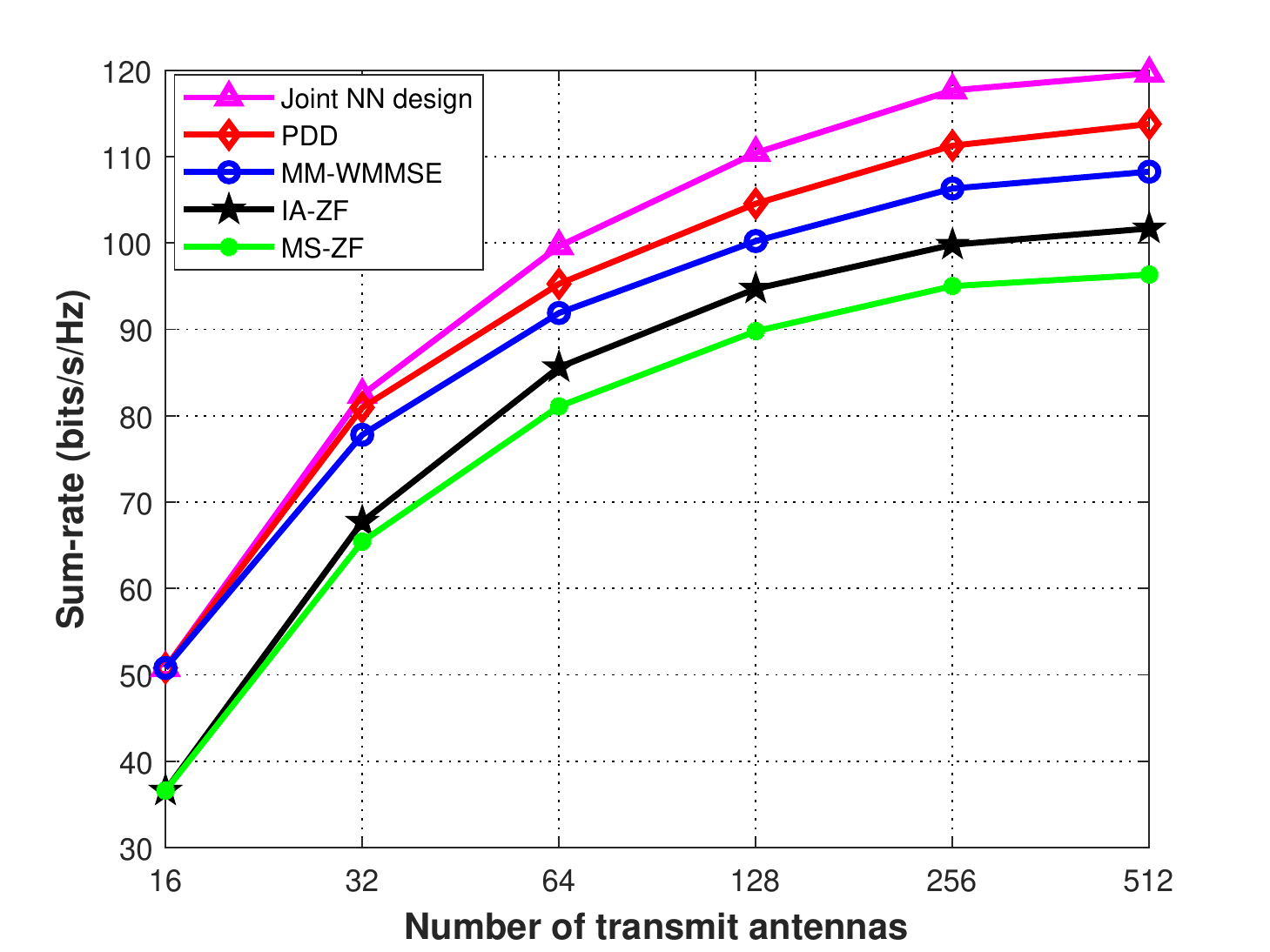}
\par\end{centering}
\caption{Achievable system sum-rate versus the number of transmit antennas $M_{s}$.}
\label{RMs}
\end{figure}

Fig. \ref{RMs} shows the achievable sum-rate when applying different numbers of transmit antennas. For this figure, we set $N_{RF}=16$ and $K=16$. The joint NN design achieves nearly the same sum-rate performance with that of the PDD and MM-WMMSE when $M_{s}=16$ since there are only $16$ beams selected for $16$ RF chains and the digital precoding designed for these three schemes are nearly the same. Similarly, the IA-ZF and MS-ZF achieve the same peformance since they both apply the ZF precoding. In addition, the gap between the former three schemes and the latter two when $M_{s}=16$ is mainly caused by the difference of WMMSE and ZF precoding. 
We can see that the performance of all schemes increases monotonically with $M_s$, and the joint NN design achieves the best sum-rate performance, followed by the PDD, MM-WMMSE, IA-ZF, and MS-ZF schemes. The gap among these schemes increases with $M_s$ since the angular resolution of the antenna increases, which leads to large distinctions among the different beams. Besides, the increasing rate slows down with $M_s$ due to the fact that the angular resolution of antenna is enough when $M_s$ is large, i.e., $M_{s}=256$.  

\begin{figure}[t]
\begin{centering}
\includegraphics[width=0.55\textwidth]{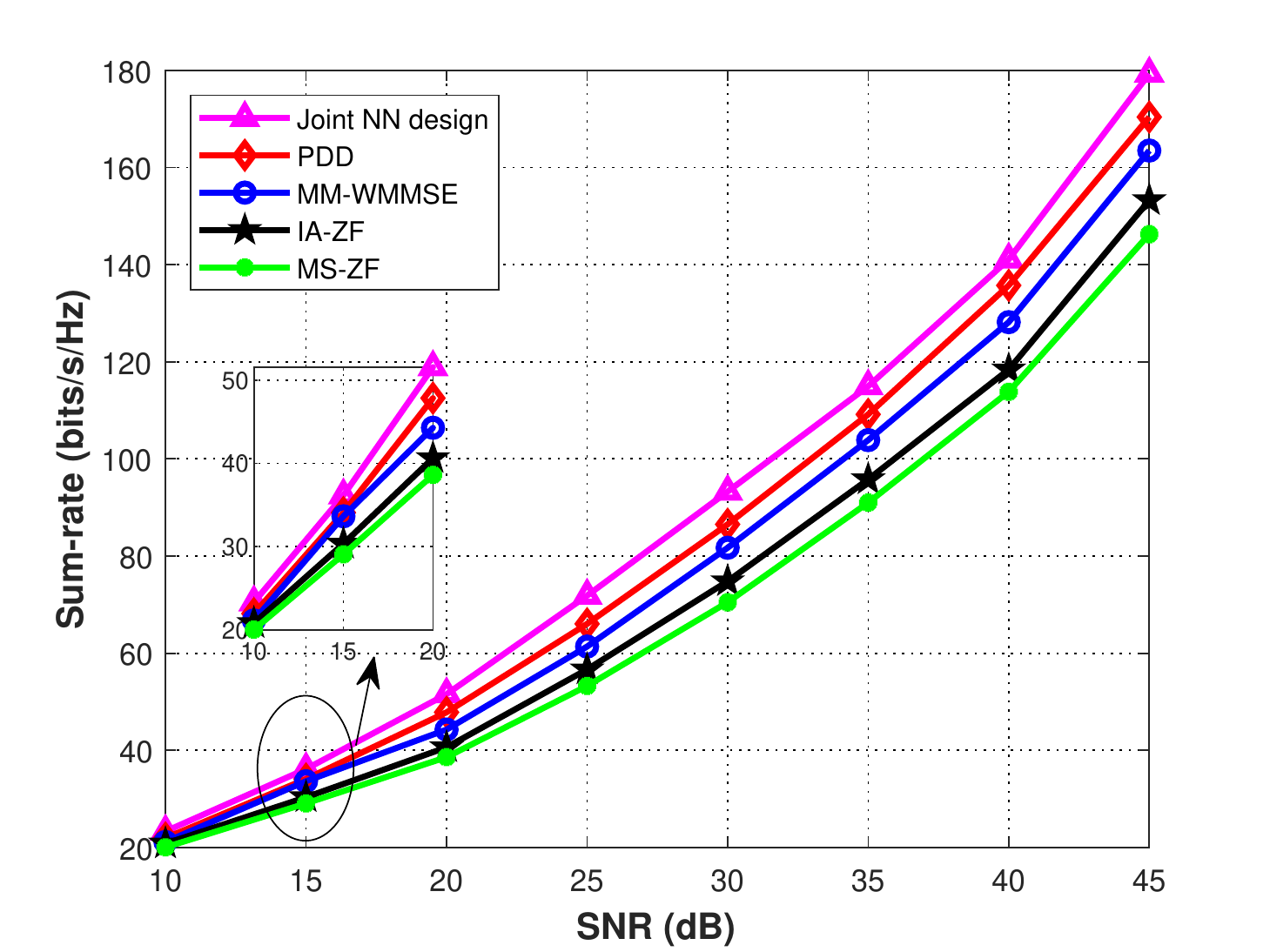}
\par\end{centering}
\caption{Achievable system sum-rate versus the SNR.}
\label{RSNR}
\end{figure}

Fig. \ref{RSNR} illustrates the achievable system sum-rate versus the SNR in different schemes. We can see that the sum-rate achieved by all algorithms increases monotonically with SNR. Besides, the joint NN design significantly outperforms the benchmarks, and the gap increases with SNR. It is mainly because in low SNR, the differences among beams are small, which leads to insignificant effects of the beam selection and digital precoding. For high SNR, e.g, $45$ dB, the joint NN design has about $6\%$ performance gain compared to the PDD, and more than $10\%$ performance gain in comparison to the other benchmarks. It verifies the advantages of the proposed joint NN design. Thus, we can conclude that the joint NN design provides an efficient and attractive solution for this problem, especially in the high SNR scenario.

\subsection{Generalization Ability}
In this subsection, we analyze the generalization ability of the proposed joint NN design. 
A jointly trained NN with system configuration $(N_{RF_{0}}, M_{s_{0}}, K_{0})$ can be straightforwardly transferred to test the samples with smaller $(N_{RF_{1}}, M_{s_{1}}, K_{1})$, rather than training a new NN. By reducing the dimension, the input channel $\mathbf{H}$ of the trained DRL-based NN has dimension $\bar{M_{s}}\times K_{0}$ ($\bar{M}_{s}< M_{s_{1}}$). To test the samples from the system configuration $(N_{RF_{1}}, M_{s_{1}}, K_{1})$, we need to input the channel with the same dimension of the training data. However, the dimension of channel in these testing data is $\bar{M_{s}}\times K_{1}$ and we need to add $(K_{0}-K_{1})$ zero column vectors. Moreover, the time step of the DRL-based NN in each episode is set to be $N_{RF_{1}}$.
As for the deep-unfolding NN, the input is the equivalent channel $\bar{\mathbf{H}}\in \mathbb{C}^{N_{RF_{0}}\times K_{0}}$ but the dimension of equivalent channel in the testing data is $N_{RF_{1}}\times K_{1}$. Thus, we need to add $(N_{RF_{0}}-N_{RF_{1}})$ zero row vectors and $(K_{0}-K_{1})$ zero column vectors to $\bar{\mathbf{H}}$.
In addition, the NN trained with a certain SNR can be directly employed to test the samples with different values of SNR since the SNR is part of the inputs of the proposed joint NN design. To improve the generalization ability of the NN with respect to the SNR, the training dataset could involve the samples with different values of SNR. 

\begin{figure}[!t]
\centering
\subfloat[]{\centering \scalebox{0.55}{\includegraphics{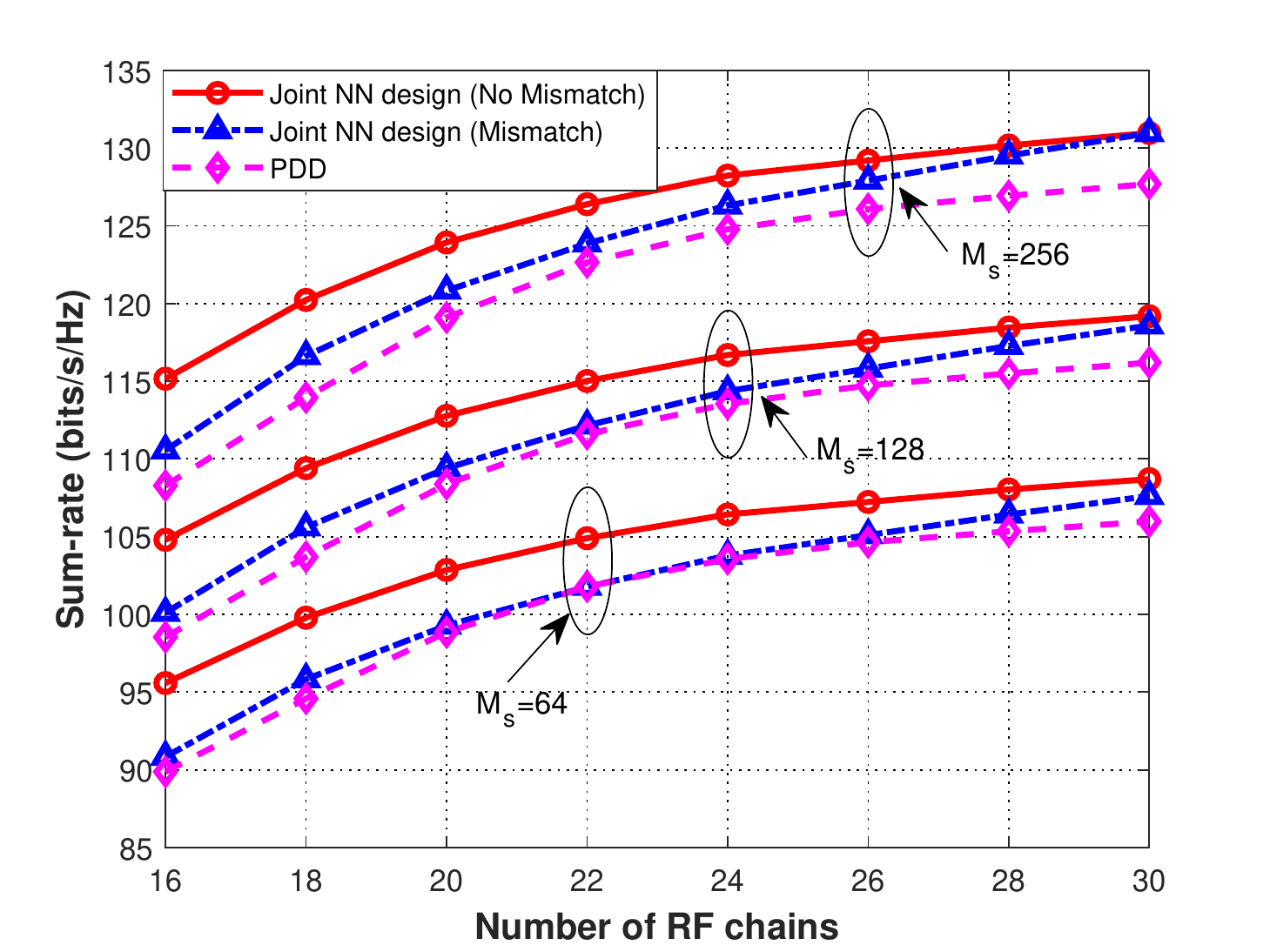}} }
\subfloat[]{\centering \scalebox{0.55}{\includegraphics{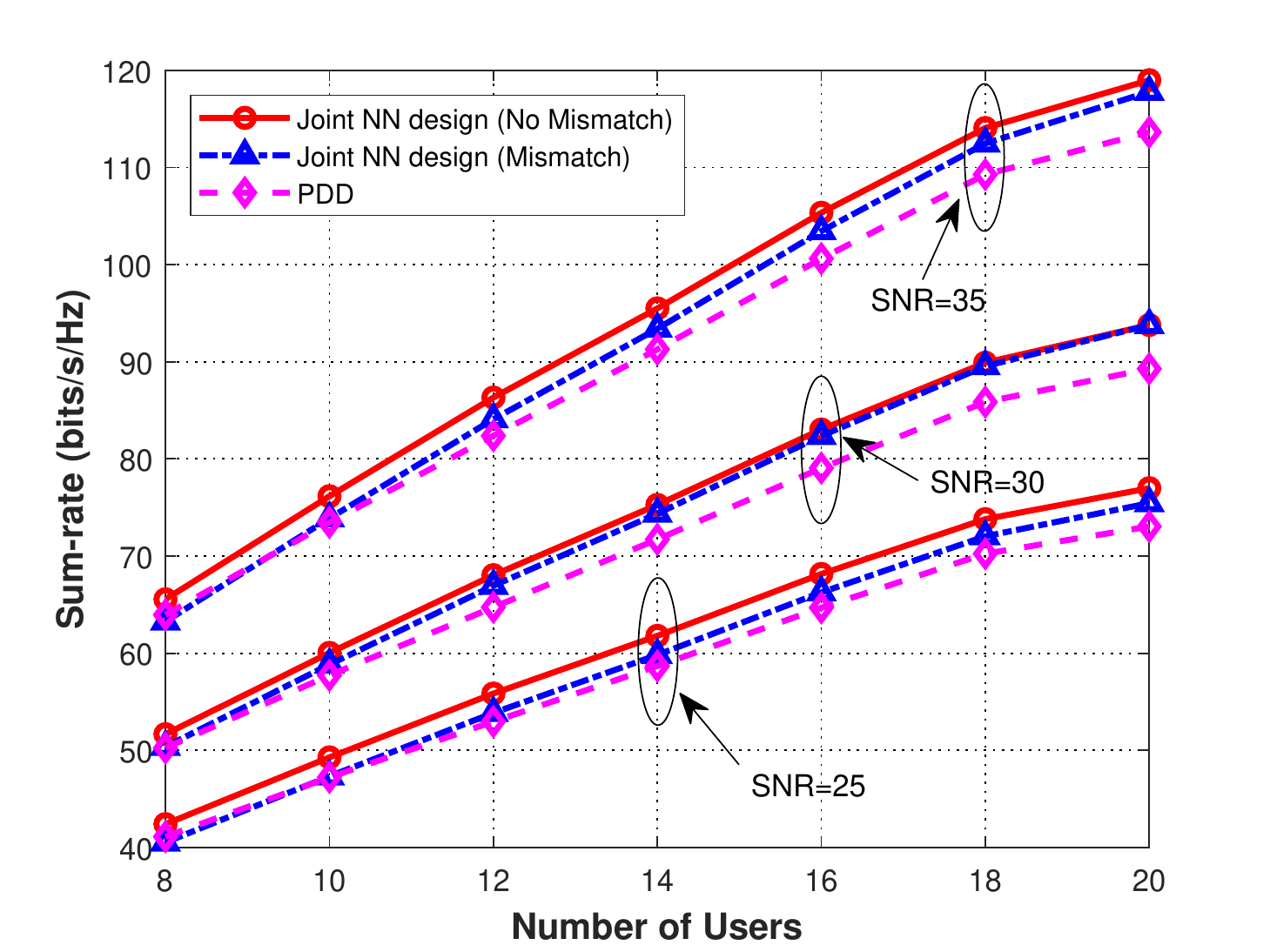}}}	
\caption{Sum-rate performance of the joint NN design with various mismatches: (a) MIMO configuration mismatch: $N_{RF}$ and $M_s$ mismatches; (b) $K$ and SNR mismatches.}
\label{General}
\end{figure}

Fig. \ref{General}(a) presents the sum-rate performance of the joint NN design with the MIMO configuration mismatch. We train the DRL-based NN and the deep-unfolding NN in the configuration of $K=16$, SNR$=40$ dB, $N_{RF}=30$, and $M_{s}=256$ and test the trained NNs in different settings of $N_{RF}$ and $M_{s}$ with fixed $K=16$ and SNR$=40$ dB. From the figure, we can see that though the NNs are employed in different MIMO configurations, there exists a small performance loss due to the mismatch. Moreover, the mismatched joint NN design still outperforms the PDD. It demonstrates the satisfactory generalization ability of the proposed joint NN for different MIMO configurations. In addition, the performance loss between the mismatched joint NN design and that without the mismatch decreases with $N_{RF}$ and $M_s$. It is mainly because the performance of these two schemes is close when the mismatch between the training and testing configurations is small.   
Fig. \ref{General}(b) presents the sum-rate performance of the joint NN design with the mismatch of SNR and $K$. We train the DRL-based NN and the deep-unfolding NN in the setting of $K=20$, SNR $=30$ dB, $N_{RF}=20$, and $M_{s}=256$ and test the trained NNs in different values of SNR and $K$ with fixed $N_{RF}=20$ and $M_{s}=256$. It is obvious that the mismatched joint NN design always outperforms the PDD even through there is a small performance loss. The sum-rate performance of the mismatched joint NN design is closer to that without the mismatch for SNR=$30$ dB compared to SNR=$25$ dB and SNR=$35$ dB. Furthermore, we can see that the performance loss decreases with $K$. The small performance loss demonstrates the satisfactory generalization ability of the proposed joint NN design for different values of $K$ and SNR.

\subsection{Fairness, Imperfect CSI, and Complexity Comparison}
\begin{figure}[t]
\begin{centering}
\includegraphics[width=0.55\textwidth]{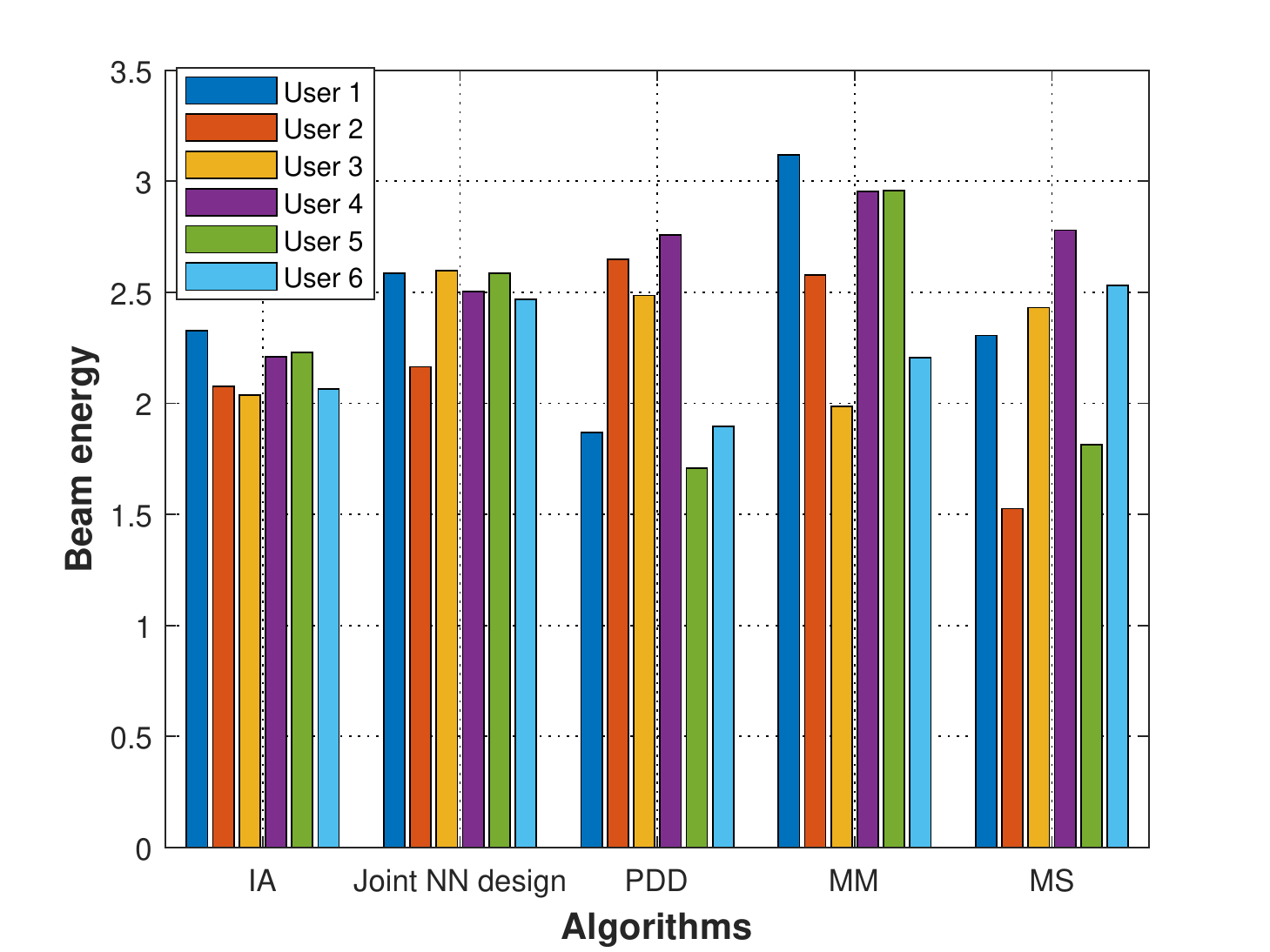}
\par\end{centering}
\caption{Selected beam energy of each user.}
\label{Fairness}
\end{figure}

Fig. \ref{Fairness} presents the selected beam energy of each user in the setting of $K=6$, SNR $=30$ dB, $N_{RF}=8$, and $M_{s}=128$, which shows the fairness among users. The selected beam energy of the $k$-th user is defined as the $l_{2}$-norm of the $k$-th column of the equivalent channel $\bar{\mathbf{H}}$, i.e., $\| \bar{\mathbf{h}}_{k} \|$. It is obvious that the selected beam energy of different users achieved by the IA and joint NN design is more balanced than that of the PDD, MM, and MS, which demonstrates that the IA and joint NN design achieves better fairness among users. It is mainly because the IA selects the beam aligned for each user and the joint NN design takes the fairness \eqref{rewardAve} into consideration.

\begin{figure}[t]
\begin{centering}
\includegraphics[width=0.55\textwidth]{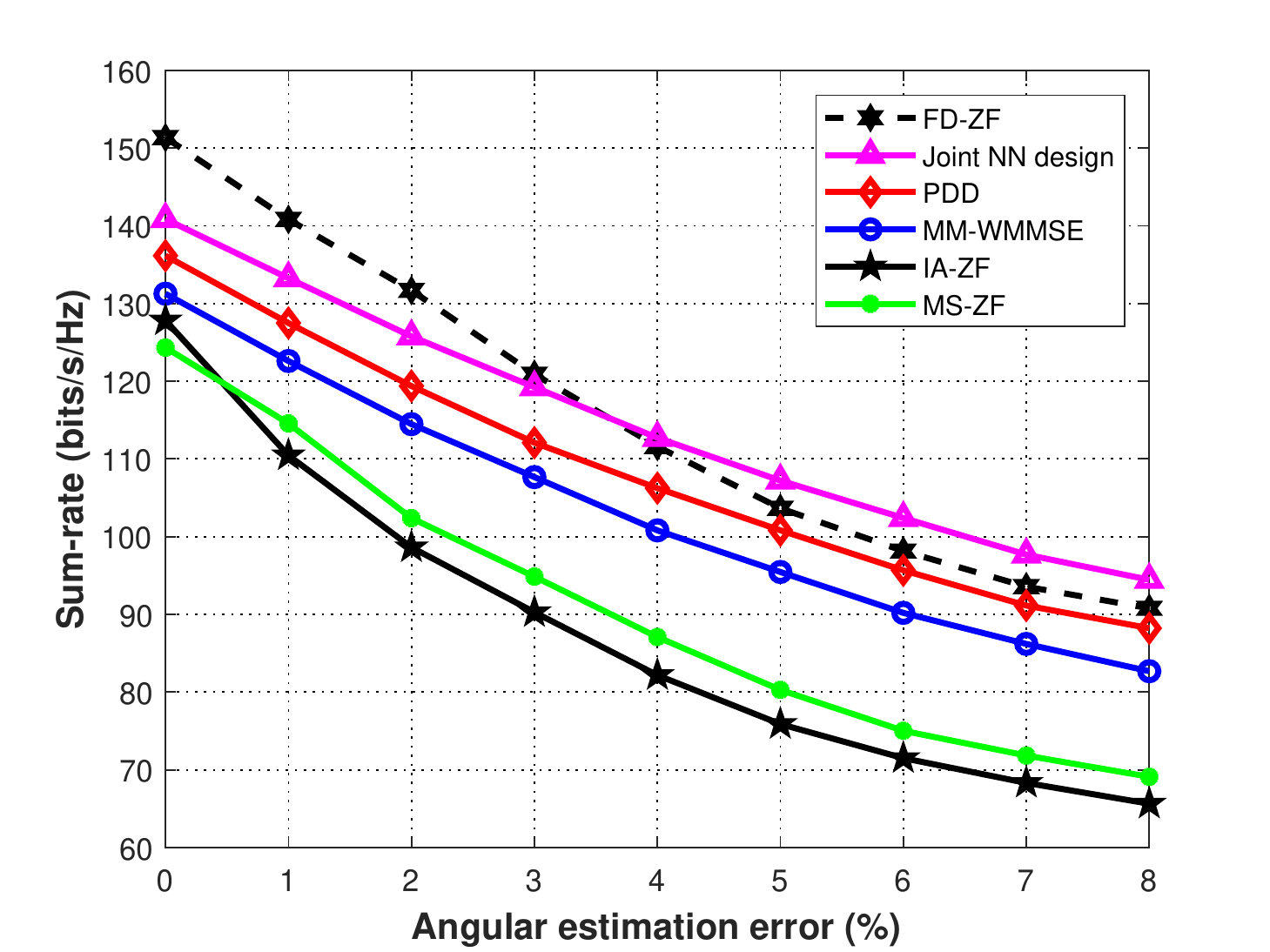}
\par\end{centering}
\caption{Achievable system sum-rate versus the channel estimation error.}
\label{ChannelErr}
\end{figure}

Fig. \ref{ChannelErr} presents the achievable system sum-rate versus the channel estimation error of the analyzed schemes. The channel estimation error is characterized by the angular estimation error defined as $\dfrac{\Delta \phi}{\frac{\pi}{M_s}}$, where $\Delta \phi$ denotes the error between the estimated angle and the actual angle of the LoS in the channel model \eqref{channel}, and $\frac{\pi}{M_s}$ is the angular resolution of the antenna. From the results, the sum-rate performance achieved by all algorithms degrades with the angular estimation error. The proposed joint NN design provides the best performance, followed by the PDD, MM-WMMSE, MS-ZF, and IA-ZF, which verifies the ability of the joint NN design to handle channel uncertainties. It is worth noting that the sum-rate achieved by the FD-ZF degrades severely and the joint NN design outperforms FD-ZF when the angular estimation error reaches $4\%$. It is mainly because: (i) The FD-ZF calculates the precoding matrix based on the original channel $\mathbf{H}$ while the other schemes apply the equivalent channel $\bar{\mathbf{H}}$ with selected beams and much lower dimension; (ii) The iterative WMMSE algorithm and deep-unfolding NN have better robustness than the ZF precoding since the ZF precoding will cause severe interference among users with channel estimation errors. Furthermore, the gap between the joint NN design and the other benchmarks escalates with the angular estimation error since the DRL-based NN and deep-unfolding NN have better robustness fed with a large number of samples in the training process.

\begin{figure}[t]
\begin{centering}
\includegraphics[width=0.55\textwidth]{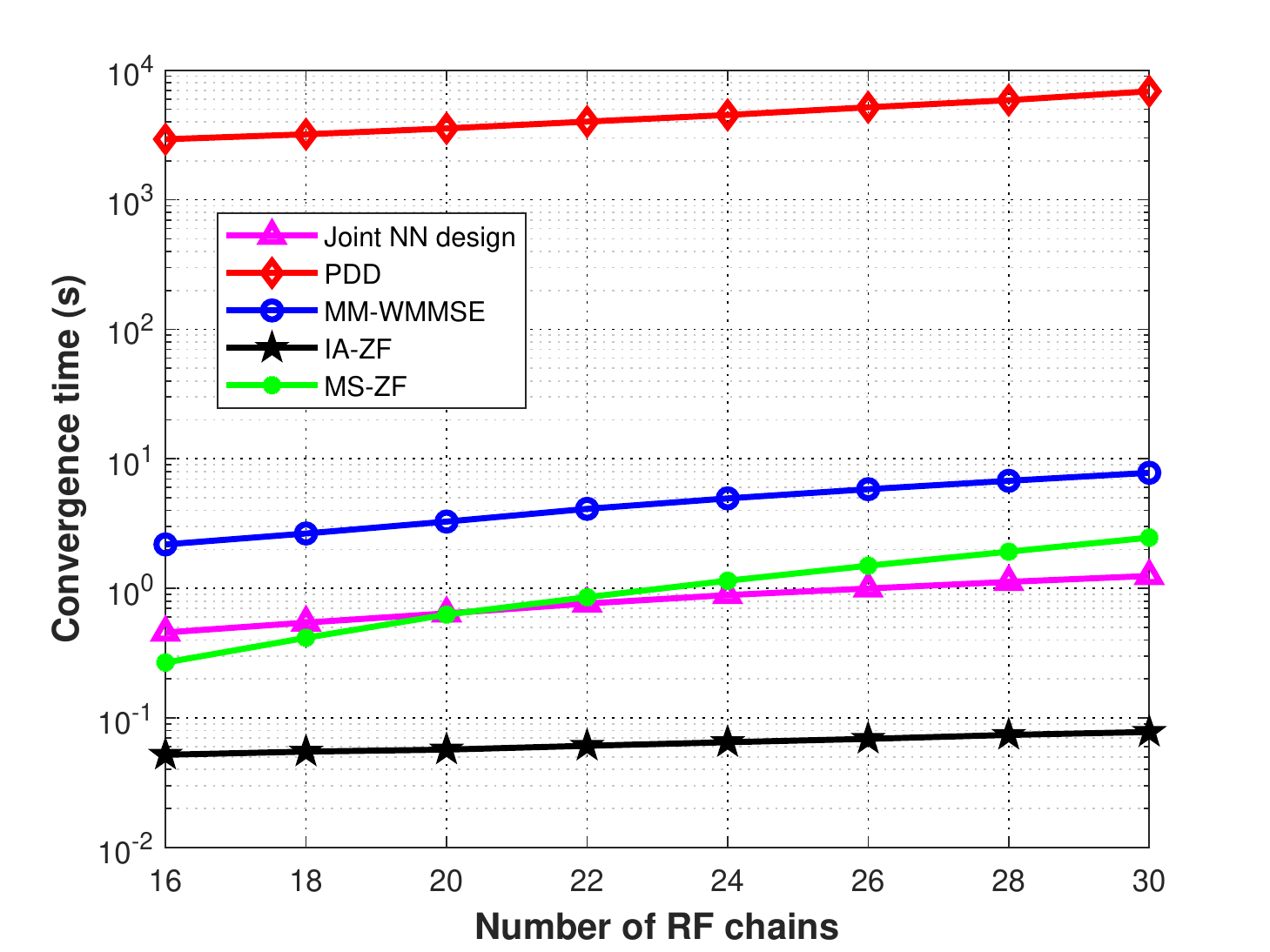}
\par\end{centering}
\caption{Convergence time versus the number of RF chains $N_{RF}$.}
\label{Time}
\end{figure}

Fig. \ref{Time} shows the convergence time versus the number of RF chains in the aforementioned schemes and we simulate the scenarios where the number of users equals to that of the RF chains, i.e., $K=N_{RF}$. The convergence time of the joint NN design is computed by averaging the inference time of $500$ samples. The convergence time increases monotonically with $N_{RF}$. Compared with the computational analysis in Table \ref{Complexity}, we can see that the PDD has the highest complexity and largest convergence time, followed by the MM-WMMSE, MS-ZF, joint NN design, and IA-ZF. Note that the convergence time of MS-ZF exceeds the joint NN design when $N_{RF}=22$. By considering the convergence time, sum-rate performance, and hardware costs, we can conclude that the proposed joint NN designed for beam selection and digital precoding achieves a good balance among the performance, cost, and computational complexity.

\section{Conclusion}
\label{Conclusion}
In this work, we investigated the joint beam selection and precoding design for mmWave MU-MIMO systems with single-sided DLA. An efficient joint NN design has been proposed to solve this challenging problem. 
Specifically, the DRL-based NN has been applied to obtain the beam selection matrix and the deep-unfolding NN has been proposed to optimize the digital precoding matrix. Simulation results showed that our proposed jointly trained NN significantly outperforms the existing iterative algorithms with reduced computational complexity and stronger robustness. 
Thus, we conclude that the proposed joint NN design can be employed as an alternative of the iterative optimization algorithm in practical systems.
The future work could generalize our proposed joint NN design into a framework to solve other challenging MINLPs in communications, where the DRL-based NN and the deep-unfolding NN can be applied to optimize the discrete variables and the continuous variables, respectively.

\bibliographystyle{IEEEtran}
\bibliography{IEEEabrv,AI_BeamSelection}

\begin{thebibliography}{10}
\providecommand{\url}[1]{#1}
\csname url@samestyle\endcsname
\providecommand{\newblock}{\relax}
\providecommand{\bibinfo}[2]{#2}
\providecommand{\BIBentrySTDinterwordspacing}{\spaceskip=0pt\relax}
\providecommand{\BIBentryALTinterwordstretchfactor}{4}
\providecommand{\BIBentryALTinterwordspacing}{\spaceskip=\fontdimen2\font plus
\BIBentryALTinterwordstretchfactor\fontdimen3\font minus
  \fontdimen4\font\relax}
\providecommand{\BIBforeignlanguage}[2]{{%
\expandafter\ifx\csname l@#1\endcsname\relax
\typeout{** WARNING: IEEEtran.bst: No hyphenation pattern has been}%
\typeout{** loaded for the language `#1'. Using the pattern for}%
\typeout{** the default language instead.}%
\else
\language=\csname l@#1\endcsname
\fi
#2}}
\providecommand{\BIBdecl}{\relax}
\BIBdecl

\bibitem{Magatech}
F.~{Boccardi}, R.~W. {Heath}, A.~{Lozano}, T.~L. {Marzetta}, and P.~{Popovski},
  ``Five disruptive technology directions for 5{G},'' \emph{IEEE Commun. Mag.},
  vol.~52, no.~2, pp. 74--80, Feb. 2014.

\bibitem{MIMO1}
T.~L. {Marzetta}, ``Noncooperative cellular wireless with unlimited numbers of
  base station antennas,'' \emph{IEEE Trans. Wireless Commun.}, vol.~9, no.~11,
  pp. 3590--3600, Nov. 2010.

\bibitem{MIMO2}
A.~{Ghosh}, T.~A. {Thomas}, M.~C. {Cudak}, R.~{Ratasuk}, P.~{Moorut}, F.~W.
  {Vook}, T.~S. {Rappaport}, G.~R. {MacCartney}, S.~{Sun}, and S.~{Nie},
  ``Millimeter-wave enhanced local area systems: A high-data-rate approach for
  future wireless networks,'' \emph{IEEE J. Sel. Areas Commun.}, vol.~32,
  no.~6, pp. 1152--1163, Jun. 2014.

\bibitem{MIMO3}
F.~{Rusek}, D.~{Persson}, B.~K. {Lau}, E.~G. {Larsson}, T.~L. {Marzetta},
  O.~{Edfors}, and F.~{Tufvesson}, ``Scaling up {MIMO}: Opportunities and
  challenges with very large arrays,'' \emph{IEEE Signal Process. Mag.},
  vol.~30, no.~1, pp. 40--60, Jan. 2013.

\bibitem{Hybrid}
O.~E. {Ayach}, S.~{Rajagopal}, S.~{Abu-Surra}, Z.~{Pi}, and R.~W. {Heath},
  ``Spatially sparse precoding in millimeter wave {MIMO} systems,'' \emph{IEEE
  Trans. Wireless Commun.}, vol.~13, no.~3, pp. 1499--1513, Mar. 2014.

\bibitem{Concept}
J.~{Brady}, N.~{Behdad}, and A.~M. {Sayeed}, ``Beamspace {MIMO} for
  millimeter-wave communications: System architecture, modeling, analysis, and
  measurements,'' \emph{IEEE Trans. Antennas Propag.}, vol.~61, no.~7, pp.
  3814--3827, Jul. 2013.

\bibitem{RZhang}
Y.~{Zeng} and R.~{Zhang}, ``Millimeter wave {MIMO} with lens antenna array: A
  new path division multiplexing paradigm,'' \emph{IEEE Trans. Commun.},
  vol.~64, no.~4, pp. 1557--1571, Apr. 2016.

\bibitem{NPhard}
H.~{Liu}, X.~{Yuan}, and Y.~{Zhang}, ``Statistical beamforming for {FDD}
  downlink massive {MIMO} via spatial information extraction and beam
  selection,'' \emph{arXiv preprint arXiv:2003.03041}, 2020.

\bibitem{NearOptim}
X.~{Gao}, L.~{Dai}, Z.~{Chen}, Z.~{Wang}, and Z.~{Zhang}, ``Near-optimal beam
  selection for beamspace mm{W}ave massive {MIMO} systems,'' \emph{IEEE Commun.
  Lett.}, vol.~20, no.~5, pp. 1054--1057, May 2016.

\bibitem{RGuo}
R.~{Guo}, Y.~{Cai}, M.~{Zhao}, Q.~{Shi}, B.~{Champagne}, and L.~{Hanzo},
  ``Joint design of beam selection and precoding matrices for mm{W}ave
  {MU-MIMO} systems relying on lens antenna arrays,'' \emph{IEEE J. Sel. Topics
  Signal Process.}, vol.~12, no.~2, pp. 313--325, May 2018.

\bibitem{GCBeam}
A.~{Sayeed} and J.~{Brady}, ``Beamspace {MIMO} for high-dimensional multiuser
  communication at millimeter-wave frequencies,'' in \emph{Proc. IEEE Global
  Telecommun. Conf. (GLOBECOM)}, Dec. 2013, pp. 3679--3684.

\bibitem{LowRF}
P.~V. {Amadori} and C.~{Masouros}, ``Low {RF}-complexity millimeter-wave
  beamspace-{MIMO} systems by beam selection,'' \emph{IEEE Trans. Commun.},
  vol.~63, no.~6, pp. 2212--2223, Jun. 2015.

\bibitem{Hanzo}
W.~{Shen}, X.~{Bu}, X.~{Gao}, C.~{Xing}, and L.~{Hanzo}, ``Beamspace precoding
  and beam selection for wideband millimeter-wave {MIMO} relying on lens
  antenna arrays,'' \emph{IEEE Trans. Signal Process.}, vol.~67, no.~24, pp.
  6301--6313, Dec. 2019.

\bibitem{Heath}
N.~J. {Myers}, A.~{Mezghani}, and R.~W. {Heath}, ``{FALP}: Fast beam alignment
  in mm{W}ave systems with low-resolution phase shifters,'' \emph{IEEE Trans.
  Commun.}, vol.~67, no.~12, pp. 8739--8753, Dec. 2019.

\bibitem{LowComp}
R.~{Pal}, A.~K. {Chaitanya}, and K.~V. {Srinivas}, ``Low-complexity beam
  selection algorithms for millimeter wave beamspace {MIMO} systems,''
  \emph{IEEE Commun. Lett.}, vol.~23, no.~4, pp. 768--771, Apr. 2019.

\bibitem{DLMaga}
Z.~{Qin}, H.~{Ye}, G.~Y. {Li}, and B.~F. {Juang}, ``Deep learning in physical
  layer communications,'' \emph{IEEE Wireless Commun.}, vol.~26, no.~2, pp.
  93--99, Apr. 2019.

\bibitem{LearnOpt}
H.~{Sun}, X.~{Chen}, Q.~{Shi}, M.~{Hong}, X.~{Fu}, and N.~D. {Sidiropoulos},
  ``Learning to optimize: Training deep neural networks for interference
  management,'' \emph{IEEE Trans. Signal Process.}, vol.~66, no.~20, pp.
  5438--5453, Oct. 2018.

\bibitem{SVMBeam}
Y.~{Long}, Z.~{Chen}, J.~{Fang}, and C.~{Tellambura}, ``Data-driven-based
  analog beam selection for hybrid beamforming under mm{W}ave channels,''
  \emph{IEEE J. Sel. Topics Signal Process.}, vol.~12, no.~2, pp. 340--352, May
  2018.

\bibitem{5GBeam}
A.~{Klautau}, P.~{Batista}, N.~{González-Prelcic}, Y.~{Wang}, and R.~W.
  {Heath}, ``{5G} {MIMO} data for machine learning: Application to
  beam-selection using deep learning,'' in \emph{2018 Information Theory and
  Applications Workshop (ITA)}, 2018, pp. 1--9.

\bibitem{MLPBeam}
C.~{Antón-Haro} and X.~{Mestre}, ``Learning and data-driven beam selection for
  mm{W}ave communications: An angle of arrival-based approach,'' \emph{IEEE
  Access}, vol.~7, pp. 20\,404--20\,415, 2019.

\bibitem{DRLresou1}
Y.~{Wei}, F.~R. {Yu}, M.~{Song}, and Z.~{Han}, ``User scheduling and resource
  allocation in {HetNets} with hybrid energy supply: An actor-critic
  reinforcement learning approach,'' \emph{IEEE Trans. Wireless Commun.},
  vol.~17, no.~1, pp. 680--692, Jan. 2018.

\bibitem{UnfoldSurvey}
A.~{Balatsoukas-Stimming} and C.~{Studer}, ``Deep unfolding for communications
  systems: A survey and some new directions,'' in \emph{2019 IEEE Int. Workshop
  on Signal Process. Systems (SiPS)}, 2019, pp. 266--271.

\bibitem{UnfoldTopic}
J.~{Chien} and C.~{Lee}, ``Deep unfolding for topic models,'' \emph{IEEE Trans.
  Pattern Anal. Mach. Intell.}, vol.~40, no.~2, pp. 318--331, 2018.

\bibitem{AMP}
M.~{Borgerding}, P.~{Schniter}, and S.~{Rangan}, ``Amp-inspired deep networks
  for sparse linear inverse problems,'' \emph{IEEE Trans. Signal Process.},
  vol.~65, no.~16, pp. 4293--4308, Aug. 2017.

\bibitem{Qiyu}
Q.~Hu, Y.~Cai, Q.~Shi, K.~Xu, G.~Yu, and Z.~Ding, ``Iterative algorithm induced
  deep-unfolding neural networks: Precoding design for multiuser {MIMO}
  systems,'' \emph{IEEE Trans. Wireless Commun.}, to appear.

\bibitem{Detection}
H.~{He}, C.~{Wen}, S.~{Jin}, and G.~Y. {Li}, ``Model-driven deep learning for
  {MIMO} detection,'' \emph{IEEE Trans. Signal Process.}, vol.~68, pp.
  1702--1715, 2020.

\bibitem{Nature}
V.~{Mnih} and \textit{et al.}, ``Human-level control through deep reinforcement
  learning,'' \emph{Nature}, vol. 518, pp. 529--233, Feb. 2015.

\bibitem{DRLoff}
L.~{Xiao}, Y.~{Li}, X.~{Huang}, and X.~{Du}, ``Cloud-based malware detection
  game for mobile devices with offloading,'' \emph{IEEE Trans. Mobile Comput.},
  vol.~16, no.~10, pp. 2742--2750, Oct. 2017.

\bibitem{DRLAccess}
S.~{Wang}, H.~{Liu}, P.~H. {Gomes}, and B.~{Krishnamachari}, ``Deep
  reinforcement learning for dynamic multichannel access in wireless
  networks,'' \emph{IEEE Trans. Cogn. Commun. Netw.}, vol.~4, no.~2, pp.
  257--265, Jun. 2018.

\bibitem{DRLradio}
Y.~{Sun}, M.~{Peng}, and S.~{Mao}, ``Deep reinforcement learning-based mode
  selection and resource management for green fog radio access networks,''
  \emph{IEEE Internet Things J.}, vol.~6, no.~2, pp. 1960--1971, Apr. 2019.

\bibitem{DRLresou2}
U.~{Challita}, L.~{Dong}, and W.~{Saad}, ``Proactive resource management for
  {LTE} in unlicensed spectrum: A deep learning perspective,'' \emph{IEEE
  Trans. Wireless Commun.}, vol.~17, no.~7, pp. 4674--4689, Jul. 2018.

\bibitem{DRLResou3}
N.~{Zhao}, Y.~{Liang}, D.~{Niyato}, Y.~{Pei}, M.~{Wu}, and Y.~{Jiang}, ``Deep
  reinforcement learning for user association and resource allocation in
  heterogeneous cellular networks,'' \emph{IEEE Trans. Wireless Commun.},
  vol.~18, no.~11, pp. 5141--5152, Nov. 2019.

\bibitem{DRLMEC}
L.~{Huang}, S.~{Bi}, and Y.~J. {Zhang}, ``Deep reinforcement learning for
  online computation offloading in wireless powered mobile-edge computing
  networks,'' \emph{IEEE Trans. Mobile Comput.}, vol.~19, no.~11, pp.
  2581--2593, Nov. 2020.

\bibitem{WMMSE}
Q.~{Shi}, M.~{Razaviyayn}, Z.~{Luo}, and C.~{He}, ``An iteratively weighted
  {MMSE} approach to distributed sum-utility maximization for a {MIMO}
  interfering broadcast channel,'' \emph{IEEE Trans. Signal Process.}, vol.~59,
  no.~9, pp. 4331--4340, Sep. 2011.

\bibitem{JointHybrid}
J.~{Tao}, J.~{Chen}, J.~{Xing}, S.~{Fu}, and J.~{Xie}, ``Autoencoder neural
  network based intelligent hybrid beamforming design for mm{W}ave massive
  {MIMO} systems,'' \emph{IEEE Trans. Cogn. Commun. Netw.}, vol.~6, no.~3, pp.
  1019--1030, Sep. 2020.

\bibitem{DDQN}
H.~{van Hasselt}, A.~{Guez}, and D.~{Silver}, ``Deep reinforcement learning
  with double {Q}-learning,'' in \emph{Proc. AAAI Conference on Artificial
  Intelligence}, 2016, pp. 2094--2100.

\end{thebibliography}

\end{document}